\documentclass[10pt]{article}
\usepackage{amssymb}
\usepackage{amsbsy}
\usepackage{verbatim}
\usepackage{epsfig}
\newcommand\DD {{\cal D}}
\def \GG{\boldsymbol \Gamma}
\def\wt{\widetilde}
\def\ds{\displaystyle}
\def\e{\mathbf e}
\def\res{\mathop{\mathrm{res}}\limits_}

\newtheorem{theorem}{Theorem}[section]
\newtheorem{remark}{Remark}[section]
\newtheorem{proposition}{Proposition}[section]
\newtheorem{example}{Example}[section]
\newtheorem{lemma}{Lemma}[section]
\newtheorem{corollary}{Corollary}[section]

\newtheorem{definition}{Definition}[section]
\makeatletter
\@addtoreset{equation}{section}
\makeatother
\def\tr{\mathrm {Tr}}
\def\le{\left}
\def\WP{\boldsymbol \wp}

\def \p{\mathbf p}
\def \r{\mathbf r}
\def\ri{\right}
\def\br{\begin{remark}}
\def\1{{\bf 1}}
\def\er{\end{remark}}
\def\bt{\begin{theorem}}

\def\et{\end{theorem}}

\def\bx{\begin{example}}
\def\ex{\end{example}}
\def\bd{\begin{definition}}
\def\ed{\end{definition}}
\def\bp{\begin{proposition}\rm}
\def\bl{\begin{lemma}\em}
\def\el{\end{lemma}}
\def\ep{\end{proposition}}
\def\be{\begin{equation}}
\def\ee{\end{equation}}
\def\bea{\begin{eqnarray}}
\def\eea{\end{eqnarray}}
\def\beaq{\begin{eqnarray}}
\def\eeaq{\end{eqnarray}}
\def \pa{\partial}
\def\B{{\mathbf B}}
\def\b{\mathfrak b}
\def\C{{\mathbb C}}
\def\O{\mathcal O}
\def\R{{\mathbb R}}
\def\N{{\mathbb N}}
\def\wr{{\mathcal R}}

\def\Z{{\mathbb Z}}

\def\one{\mathbf 1}

\begin{document}
\def\L{\Lambda}
\begin{flushright}
CRM-3182 (2005)
\end{flushright}
\vspace{0.2cm}
\begin{center}
\begin{Large}
\textbf{Biorthogonal Laurent polynomials,  T\"oplitz determinants, minimal Toda orbits and  isomonodromic tau functions}
\end{Large}\\
\vspace{1.0cm}
{\bf M. Bertola\footnote{Work supported in part by the Natural
    Sciences and Engineering Research Council of Canada (NSERC),
    Grant. No. 261229-03 and by the Fonds FCAR du
    Qu\'ebec.}\footnote{e-mail: bertola@mathstat.concordia.ca},
  M. Gekhtman\footnote{Work supported in part by the National Science
    Foundation, Grant No. DMS-0400484.}\footnote{e-mail: gekhtman.1@nd.edu} }\par 
\bigskip
\bigskip
\bigskip
%%%%%%%%%%%%%%%%%%%%%%%%%%%%%%%%%%  Abstract %%%%%%%%%%%%%%%%%%%%%%%%%%%%%%%
{\bf Abstract}\\
\end{center}
 We consider the class of biorthogonal polynomials that are used to
 solve  the inverse spectral problem associated to  elementary
 co-adjoint orbits of the Borel group of upper triangular matrices;
 these orbits are the  phase space of
 generalized integrable lattices of Toda type. Such polynomials
 naturally interpolate between the theory of orthogonal polynomials on the
 line and orthogonal polynomials on the unit circle and tie together
 the theory of  Toda, relativistic Toda, Ablowitz-Ladik and Volterra
 lattices. 
 We establish corresponding  Christoffel-Darboux formul\ae . For all these classes of polynomials a $2\times 2$ system of
 Differential-Difference-Deformation equations is analyzed in the most
 general setting of pseudo measures with arbitrary rational
 logarithmic derivative. They provide particular classes of
 isomonodromic deformations of rational connections on the Riemann
 sphere. The corresponding isomonodromic tau function is explicitly
 related to the shifted T\"oplitz determinants of the moments of the
 pseudo-measure.
 In particular the results imply that any (shifted) T\"oplitz (H\"ankel) determinant of a symbol (measure) with arbitrary rational logarithmic derivative is an isomonodromic tau function.

%%%%%%%%%%%%%%%%%%%%%%%%%%%%%%% 1. Introduction   %%%%%%%%%%%%%%%%%%%%%%%%

\tableofcontents
\section{Introduction}
The connection between orthogonal polynomials on the line and Toda
lattices is rather well known \cite{Bere}, as well as the relations to
the KP hierarchy \cite{AvM1}. Dynamical variables of the Toda lattice are arranged
into a tri-diagonal Lax matrix, that can be viewed as  a recurrence matrix for a system of orthogonal polynomials.
In the (semi)finite case, the evolution of the corresponding measure provides a linearization of the Toda flows.
More generally, one can
set-up (in)finite-dimensional Hamiltonian systems on $\R^{2n}\ (n\leq \infty) $
with Hamiltonians \bea H_I (\underline q,\underline p) = \frac 1 2
\sum_{i=1}^n {p_i}^2 + \sum_{i\not\in I} p_i {\rm e}^{q_{i+1}-q_i}
+ \sum_{j=1}^{|I|} {\rm
  e}^{q_{i_{j+1}} -q_{i_j}}\\
I:=\{i_1<i_2<\dots<i_k\}\ .
\eea
As it is noted in \cite{FG2} such family of Hamiltonians (labeled by
the multi-index $I$) contains  integrable lattice hierarchies of
Toda, relativistic Toda, Volterra and Ablowitz-Ladik type. These
integrable Hamiltonian systems have a Lax representation with Lax
operator given as a $n\times n$ lower Hessenberg matrix which we denote
by $Q$ (in \cite{FG1,FG2} it was denoted by $X$), belonging to
a certain ``elementary'' $(2n-2)$-dimensional co-adjoint orbit of the solvable group of upper
triangular matrices. These systems are linearized by the Moser map
\bea
Q\mapsto \mathcal W(z;Q):= \le(z\1 - Q \ri )^{-1}_{11} = \sum_{j=0}^\infty \frac {\hat \mu_j(Q)} {z^{j+1}}\\
\hat \mu_j(Q) = {Q^j}_{11}
\eea
In the case of infinite lattices these expressions take on a formal
meaning in terms of power series but the analysis is unchanged.\\
The  {\em moments} $\hat\mu_j$ of $Q$ define a {\em normalized} moment
functional $\mathcal L$ and the reconstruction of $Q$ from its moments
(the ``inverse moment problem'') can be accomplished by constructing a
suitable sequence of biorthogonal (Laurent) polynomials $\{r_i,p_i
\}_{i\in \N}$
$$
\mathcal L(r_i p_j) = \delta_{ij}\ ,
$$
where $p_i$'s are polynomials in $x$ of degree $i$ while $r_i$'s
are, in general, polynomials in $x$ and $x^{-1}$.  The (infinite) Lax
operator $Q$ corresponding to the chosen orbit is then reconstructed
by \cite{FG2}
$$
Q_{ij} = \mathcal L (r_i\,x\,p_j)\ .
$$
Explicit formul\ae\ for these biorthogonal polynomials in terms of
shifted T\"oplitz determinants can be found in \cite{FG1,FG2} and will be recalled here in due time.
Vice versa, one could assign an arbitrary (generic) moment functional $\mathcal L:\C[z,z^{-1}]\to \C$, a multi-index $I$ and then reconstruct the Lax operator $Q_I$ (i.e. view the Lax operator as a function of $\mathcal L$ rather than the other way around)
$$
\mathcal L\mapsto Q_I(\mathcal L).
$$
From this point of view, the linearization of the (infinite) Hamiltonian hierarchy is accomplished simply by
 \be
 \mathcal L_{\bf t}(\bullet) = \mathcal L ({\rm e}^{\sum_i 1/ i t_i z^i} \bullet)\ ,
 \ee
 where the series may  have to be understood formally.
 This procedure displays the common nature of all the above-mentioned integrable lattices, inasmuch as  the linearizing space is always the same (the space of moment functionals) and what changes from one lattice to another is only the orbit, namely the map $Q_I$.

 Finite dimensional systems(of dimension $2n-2$) on an elementary orbit $\mathfrak Q_I$ correspond to those moment-functionals for which  certain shifted T\"oplitz determinants of size $\leq n$ do not vanish whereas all larger ones do. In such cases, the tau function of the hierarchy defined by the (closed) differential
 \be
 {\rm d}\ln \tau = \sum_{J=1}^n \frac 1 J \tr_n(Q_I) {\rm d}t_J
 \ee
 and coincides with the largest non-vanishing (shifted) T\"oplitz determinant.

 One of the main purposes of this paper is to connect this determinant to a different notion of ``tau'' function, namely the one introduced by Jimbo, Miwa and Ueno in \cite{JMU,JMII}.
  It was shown in \cite{BEH, semiiso} that the H\"ankel determinants of an arbitrary (generic) ``semiclassical'' moment functional on the space of polynomials can be identified with the isomonodromic tau function introduced by our Japanese colleagues. \\
  Similarly, it was shown in \cite{ITW} that T\"oplitz determinants of a particular class of symbols on the unit circle are also identifiable with the same kind of isomonodromic tau functions.

  These two apparently distinct situations are in fact the two ends of a ``continuous'' spectrum of situations: in fact the case of H\"ankel determinants is dealt with in the setting of (generalized) ordinary orthogonal polynomials, whereas that of T\"oplitz determinants uses orthogonal polynomials on the unit circle; in this latter situation one considers polynomials $p_i(z)$ orthogonal in the usual $L^2(S^1,{\rm d}\mu)$ sense
\be
 \int_{S^1} p_j(z)\overline{p_k(z)} {\rm d}\mu(z) = \delta_{jk}\ .
\ee
Here  one defines $r_j(z) = \overline{p_j}(z^{-1})$ and the orthogonality is recast into
\be
\mathcal L(r_j p_k) = \delta_{jk}\ ,
\ee
where - in this special case -
\be
\mathcal L:\C[z,z^{-1}]\to \C\ ;\ \ \mathcal L(z^j) = \int_{S^1} z^j{\rm d}\mu(z)\ .
\ee
We see that we can regard the case of orthogonal polynomials on the
circle as a special case of biorthogonal Laurent polynomials with
respect to a moment functional satisfying the reality condition $\mu_k
= \overline {\mu_{-k}}$.

According to the previous description of integrable lattices, the two
situations correspond to two different elementary orbits and hence we
should be able to treat them on a common ground, together with all the
other lattices associated with the orbits $\mathfrak Q_I$.
Indeed, we will show that this is the case and that for the class of
moment functionals of the semiclassical type introduced in
\cite{semiiso} all the shifted T\"oplitz determinants which arise as
tau functions of the corresponding integrable lattices are also
isomonodromic tau functions for a rational $2\times 2$ connection on
$\C ^1$ which will be explicitly constructed in the paper.

The approach to this problem follows the strategy used in
\cite{semiiso} rather the one in \cite{ITW}; in the course of our
analysis we will obtain generalized Christoffel-Darboux identities
which naturally interpolate between the ordinary CD identity for
orthogonal polynomials on the line and the one for orthogonal
polynomials on the unit circle.

Moreover we will show that the T\"oplitz and H\"ankel determinants of
 the same size for one such moment functional are connected by a
 sequence of elementary Schlesinger transformations, at each step of
 which we obtain tau functions associated to interpolating
 orbits; in figurative terms, we show that the papers \cite{ITW} (see
 Example \ref{its}) and \cite{semiiso} are connected by a Schlesinger
 transformation (when specializing the semiclassical measure to the one
 relevant for \cite{ITW}) and that ``neighboring'' elementary
 co-adjoint orbits are also connected by an elementary Schlesinger
 transformation.

{\bf Acknowledgements}.
The authors thank John Harnad for stimulating discussions. M. G. is grateful
to Laboratoire de Physique Mathematique,  Centre de Recherches Mathematique
and the Concordia University for their hospitality during his visit to Montreal, where the work on this project has started. During the later stages of preparation of the manuscript he also enjoyed hospitality
of Institut des Hautes \'Etudes Scientifiques.

\section{Setting}
We start in the most general and abstract setting, without any
reference to a (pseudo) measure. We consider an arbitrary moment
functional
\be
\mathcal L:\C[z,z^{-1}] \to \C
\ee
on the space  polynomials in $z$ and $z^{-1}$ and denote its moments with $\mu_j =
\mathcal L(z^j)\ ,\ j\in \Z$.   We introduce the following {\em shifted T\"oplitz}
determinants and polynomials
\bea
\Delta_n^\ell =\det\pmatrix {\mu_{\ell} & \mu_{\ell+1} & \cdots &\mu_{\ell+n-1}\cr
\mu_{\ell-1} & \mu_\ell & \cdots & \mu_{\ell+n-2}\cr
 & \ddots & \ddots & \cr
\mu_{\ell-n+1} & \mu_{\ell-n+2} & \cdots & \mu_{\ell}}\\
\Delta_0^\ell\equiv 1 \ ,\ \ \Delta_{-n}^{\ell} \equiv 0\nonumber \\
\wp_n^\ell(x):= \det\pmatrix {\mu_{\ell} & \mu_{\ell+1} & \cdots &\mu_{\ell+n}\cr
\mu_{\ell-1} & \mu_\ell & \cdots & \mu_{\ell+n-1}\cr
 & \ddots & \ddots & \cr
\mu_{\ell-n+1} & \mu_{\ell-n+2} &\cdots & \mu_{\ell+1}\cr
1& x & \cdots & x^n}
\eea

Using some classical identities for determinants we can derive
recurrence relations for the shifts $n\to n+1$ and $\ell\to \ell+1$
for the above polynomials.
We first need the following
\bp
For any $(n+1)\times(n+1)$ matrix $A$ the following determinant
identity holds true (Jacobi identity)
\be
A^{1..n}_{1..n} A^{2..n+1}_{2..n+1} -
A^{2..n+1}_{1..n}A^{1..n}_{2..n+1} =
A_{1..n+1}^{1..n+1}A_{2..n}^{2..n}\ ,\label{jac1}
\ee
where the sub/super-script ranges denote the rows/columns of the
submatrix we are computing the determinant of.
As a corollary, for any $(n+1)\times(n+2)$ matrix $B$ we have
\be
B_{2..n+2}^{1..n+1} B_{1..n}^{1..n}\,\, +\,\,
B_{1..n+1}^{1..n+1}B_{2..n,n+2}^{1..n} = B_{1..n,n+2}^{1..n+1}
B_{2..n+1}^{1..n}\label{jac2}
\ee
which can be obtained from (\ref{jac1}) by adjoining an appropriate
row.
\ep
Using (\ref{jac1}) on the determinant defining $\wp_n^\ell$ we find
\be
x\Delta_n^\ell \wp_{n-1}^\ell - \Delta_n^{\ell+1} \wp_{n-1}^{\ell-1} =
\Delta _{n-1}^\ell \wp_n^\ell\ .
\ee
Applying (\ref{jac2}) to the determinant defining $\wp_n^\ell$
adjoined of the next row of moments on the top we find
\bea
&& \wp_n^{\ell-1} \Delta_n^\ell + \Delta_{n+1}^\ell \wp_{n-1}^{\ell-1} =
\wp_n^\ell \Delta_n^{\ell-1}\label{uno}\\
&& \wp_n^{\ell-1}\Delta_n^{\ell+1} + x \Delta_{n+1}^{\ell}
\wp_{n-1}^\ell = \wp _n^\ell \Delta_n^\ell\label{due}\\
&& x \Delta_n^\ell \wp_{n-1}^\ell -\Delta_n^{\ell+1} \wp_{n-1}^{\ell-1}
= \wp_n^\ell \Delta_{n-1}^\ell\label{tre}
\eea
We now use these identities to express
$\WP_n^\ell:=[\wp_n^\ell,\wp_{n-1}^{\ell-1}]$ in terms of $\WP_{n-1}^\ell=[\wp_{n-1}^\ell,\wp_{n-2}^{\ell-1}]$
\bea
&&\le[\matrix{\wp_n^\ell\cr \wp_{n-1}^{\ell-1} }\ri]=
\le[
\begin{array}{cc}
\ds \frac {x\Delta_n^\ell}{\Delta_{n-1}^\ell} - \frac
      {\Delta_n^{\ell+1}\Delta_{n-1}^{\ell-1}} {(\Delta_{n-1}^\ell)^2}
&
\ds \frac{\Delta_{n}^{\ell+1} \Delta_n^\ell}{(\Delta_{n-1}^{\ell})^2}
\\[15pt]
\ds \frac {\Delta_{n-1}^{\ell-1}}{\Delta_{n-1}^\ell}
&
\ds -\frac {\Delta_n^\ell}{\Delta_{n-1}^\ell}
\end{array}
\ri]\le[\matrix{\wp_{n-1}^\ell\cr \wp_{n-2}^{\ell-1} }\ri]\label{circle}\\
&&\WP_{n}^\ell = \mathcal C_n^\ell \WP_{n-1}^\ell\\
&&\det\mathcal C_n^\ell = -x\frac{(\Delta_{n}^\ell)^2}
  {(\Delta_{n-1}^\ell)^2} \ \
 \hspace{5cm}  \hbox{\bf Circle Case}  \\
&& \mathbf j \mathcal C_n^\ell(x)^{-1} \mathcal C_n^\ell(y) -\mathbf
  j= \le(1-\frac y x \ri)\le[
\begin{array}{cc}
\ds\frac{\Delta_{n-1}^{\ell-1}}{\Delta_n^\ell } & 0\\[15pt]
\ds -1 & 0
\end{array}
\ri]
\eea
where
\be
\mathbf j := \le[\begin{array}{cc}0 & 1\\ -1 & 0\end{array}\ri]\ .
\ee
We have named this  the ``circle case'' because this sort of recursion is
relevant for orthogonal polynomials on the unit circle.
We next derive a recursion in $\ell$
\bea
&& \le[\matrix{\wp_n^\ell\cr \wp_{n-1}^{\ell-1} }\ri]=
\le[
\begin{array}{cc}
\ds \frac {\Delta_n^\ell}{\Delta_{n}^{\ell-1}} + \frac
      {\Delta_{n+1}^{\ell}\Delta_{n-1}^{\ell-1}} {x(\Delta_{n}^{\ell-1})^2}
&
\ds \frac{\Delta_{n+1}^{\ell} \Delta_n^\ell}{x (\Delta_{n}^{\ell-1})^2}
\\[15pt]
\ds \frac {\Delta_{n-1}^{\ell-1}}{x\Delta_{n}^{\ell-1}}
&
\ds \frac {\Delta_n^\ell}{x\Delta_{n}^{\ell-1}}
\end{array}
\ri]\le[\matrix{\wp_{n}^{\ell-1}\cr \wp_{n-1}^{\ell-2} }\ri] \label{circleline}\\
&& \WP_n^\ell = \mathcal T_n^\ell \WP_n^{\ell-1}\\
&&\det \mathcal T_n^{\ell} = \frac 1 x \frac{(\Delta_{n}^\ell)^2}
  {(\Delta_{n}^{\ell-1})^2}\ \ \hspace{5cm}
  \hbox{\bf Circle to Line Transform} \\
&& \mathbf j \mathcal T_n^\ell(x)^{-1} \mathcal T_n^\ell(y) -\mathbf j
=
\le(1-\frac x y \ri)\le[
\begin{array}{cc}
\ds\frac{\Delta_{n-1}^{\ell-1}}{\Delta_n^\ell } & 1\\[15pt]
\ds 0 & 0
\end{array}
\ri]
\eea
The name  ``circle-to-line'' refers to the fact that this recursion relation interpolates
between the previous ``circle'' case and the next one, which will
be named the ``line'' case.
Indeed, composing these two we can express
$\WP_{n}^{\ell} = [\wp_n^\ell,\wp_{n-1}^{\ell-1}]$ in terms of $\WP_{n-1}^{\ell-1}=[\wp_{n-1}^{\ell-1},\wp_{n-2}^{\ell-2}]$
\bea
&& \le[\matrix{\wp_n^\ell\cr \wp_{n-1}^{\ell-1} }\ri]=
\le[
\begin{array}{cc}
\ds \frac {\Delta_n^\ell}{\Delta_{n-1}^{\ell-1}}\le(x +
  \frac{\Delta_{n-2}^{\ell-1}\Delta_n^\ell - \Delta_n^{\ell+1}
  \Delta_{n-1}^{\ell-1}}{\Delta_{n-1}^\ell \Delta_{n-1}^{\ell-1}}\ri)
&
\ds \frac {(\Delta_n^\ell)^2}{(\Delta_{n-1}^{\ell-1})^2}
\\[15pt]
\ds 1
&
\ds 0
\end{array}
\ri]\le[\matrix{\wp_{n-1}^{\ell-1}\cr \wp_{n-2}^{\ell-2} }\ri]\label{line}\\
&&  \WP_n^\ell = \mathcal L_n^\ell \WP_{n-1}^{\ell-1} \hspace{7cm}
  \hbox{\bf  Line case} \\
&&\det \mathcal L_n^{\ell} = - \frac{(\Delta_{n}^\ell)^2}
  {(\Delta_{n-1}^{\ell-1})^2}\\
&& \mathbf j \mathcal L_n^\ell(x)^{-1} \mathcal L_n^\ell(y) -\mathbf
  j=
\le(x-y\ri)\le[
\begin{array}{cc}
\ds\frac{\Delta_{n-1}^{\ell-1}}{\Delta_n^\ell } & 0\\[15pt]
\ds 0 & 0
\end{array}
\ri]
\eea
This recursion is called ``line'' case because it is the relevant
recursion relation for ordinary orthogonal polynomials on the line.
\subsection{Second--kind polynomials}
Let us define the following {\em second-kind} polynomials
\bea
\wr_n^\ell(x) &=& \mathcal L_z\le(\frac
   {\wp_{n}^\ell(x)-\wp_n^\ell(z)}{x-z} \ri)
\eea
The three types of recursion (\ref{circle}, \ref{circleline},
\ref{line}) involve at most a multiplication or
division by $x$ and have otherwise constant coefficients (in $x$):
moreover we find
\bea
x\wr_n^\ell(x) &=& \mathcal L_z\le(\frac
{x\wp_{n}^\ell(x)-z\wp_n^\ell(z)}{x-z} \ri) - \mathcal
L_z(\wp_n^\ell(z))\\
x^{-1}\wr_n^\ell(x) &=& \mathcal L_z\le(\frac
{x^{-1}\wp_{n}^\ell(x)-z^{-1}\wp_n^\ell(z)}{x-z} \ri) -\frac 1x  \mathcal
L_z(z^{-1}\wp_n^\ell(z))
\eea
The last terms in these identities vanish because of the determinant
structure of $\wp_n^\ell$, provided that $n\geq 1$ and
$0\leq \ell \leq n-1$ for the first case and $-1\leq \ell \leq n-2$
for the second case.
From this observation we find that these auxiliary sequences of
polynomials satisfy the same recurrence relations in the following ranges
\bea
\le[ \matrix {\wr_n^\ell\cr \wr_{n-1}^{\ell-1}}  \ri] = \le\{
\begin{array}{lll}
\mathcal L_{n}^{\ell} \le[ \matrix {\wr_{n-1}^{\ell-1}\cr
    \wr_{n-2}^{\ell-2} } \ri]  & & 1\leq \ell\leq n-1\\[10pt]
\mathcal C_{n}^{\ell} \le[ \matrix {\wr_{n-1}^{\ell}\cr
    \wr_{n-2}^{\ell-1} } \ri]  & & 0\leq \ell\leq n-2 \\[10pt]
\mathcal T_{n}^{\ell} \le[ \matrix {\wr_{n-1}^{\ell-1}\cr
    \wr_{n-2}^{\ell-2} } \ri]  & & 0\leq \ell\leq n-1\\[10pt]
\end{array}
\ri.
\eea

\section{Christoffel-Darboux formul\ae}
Consider $(n,l)\in \N\times\N$ and choose an arbitrary path
starting at the origin of the following type
\be
\{(n_k,\ell_k),\ k=0,1,\dots\ ,(n_0,\ell_0) = (0,0), (n_1,\ell_1) = (1,0)\}
\ee
and such that the possible subsequent  moves are right, up or
up-right. For the move $(n_{k-1},\ell_{k-1})\mapsto(n_k\ell_k)$ we
introduce the { \bf transfer matrices} following an idea of
\cite{goli} used for orthogonal polynomials on the circle
\be
T_k(x):= \le\{
\begin{array}{cl}
\mathcal C_{n_k}^{\ell_k} & \hbox { if the move is right ({\bf circle move}) }\\
\mathcal T_{n_k}^{\ell_k} & \hbox { if the move is up ({\bf
    circle-to-line  move}) }\\
\mathcal L_{n_k}^{\ell_k} & \hbox { if the move is up-right  ({\bf line move})}
\end{array}\ri. \label{transfer}
\ee
Using these transfer matrices we define the two dual auxiliary sequences of matrices as follows
\bea
&& \Xi_k(x) = T_k(x)\Xi_{k-1}(x) \\
&& \Xi_k^\star(x) =\frac 1 {\det T_k(x)}\Xi_{k-1}^\star(x) T_k^t(x)\
\qquad\\
&& \Xi_0^\star = \Xi_0^t.
\eea
This definition in particular implies that
\bea
\Xi_k^\star = \frac 1{\prod_{j=1}^{k} \det T_j} \Xi_k^t\ .
\eea
The choice of the initial conditions for the auxiliary sequences is arbitrary but it is convenient
to choose $\Xi_0$ in such a way that the first column of $\Xi_n$ will contain
$\wp_n^\ell$ and $\wp_{n-1}^{\ell-1}$ and the second column the
corresponding second kind polynomials. Since the matrices constructed
with the polynomials $\wp_{n_k}^{\ell_k}$ and the second kind
polynomials already satisfy the same recursion relation for $k\geq
1$, it is sufficient to impose the same initial conditions with the following choice (recall that the first move is
always a circle-move)
\bea
\Xi_0 = (C_{n_1}^{\ell_1})^{-1} \le[\matrix{ \wp_{n_1}^{\ell_1} &
    \wr_{n_1}^{\ell_1} \cr\wp_{n_1-1}^{\ell_1-1} &
    \wr_{n_1-1}^{\ell_1-1}}\ri] =
\frac 1 {\mu_0^2 x } \le[\matrix{ \mu_0 & \mu_1\mu_0\cr 1 & \mu_1-\mu_0 x}\ri]  \le[\matrix{
   \mu_0 x-\mu_1 & {\mu_0}^2  \cr 1  & 0}\ri] = \le[\begin{array}{cc}
1 & \frac {\mu_0} x \\[10pt]
 0 &    \frac 1 x \end{array}\ri]
\eea

Recall that for any $2\times 2$ matrix we have $A^t = \det(A) \mathbf
j A^{-1}\mathbf j^{-1}$.
We now compute
\bea
\Xi_{k}^\star(x)\,\mathbf j \,\Xi_{k}(y)& =& \frac 1 {\det T_{k}(x)}
\Xi_{k-1}^\star(x)\,T_{k}^t(x)\,\mathbf j \,T_{k}(y)\,\Xi_{k-1}(y)
=\nonumber \\&=&
\Xi_{k-1}^\star(x)\,\mathbf
j\,T_{k}^{-1}(x)\,T_k(y)\,\Xi_{k-1}(y)=\nonumber \\
&=&\Xi_{k-1}^\star(x)\,\mathbf j \,\Xi_{k-1}(y) +
\Xi_{k-1}^\star(x)\,\le(\mathbf j\,T_k^{-1}(x)\,T_k(y)-\mathbf j\ri)\,\Xi_{k-1}(y)
\label{ads}
\eea
Let us define $\dot\ell_k:= \ell_k-\ell_{k-1}$ and $\dot n_k :=
n_k-n_{k-1}$. Then the three formul\ae\
(\ref{circle},\ref{circleline},\ref{line}) can be uniformly written
\bea
&&\mathbf j\,T_k^{-1}(x)\,T_k(y)-\mathbf j =(-1)^{1-\dot n_k} \le(\frac 1 y -\frac 1
x\ri)x^{\dot \ell_k}  y^{\dot n_k}
\le[\begin{array}{cc}
\ds \frac{\Delta_{n_k-1}^{\ell_k-1}}{\Delta_{n_k}^{\ell_k}} &
1-\dot n_k \\[15pt]
 \dot \ell_k-1 & 0
\end{array}\ri]\\
&& \det T_k(x) = (-1)^{\dot n_k} x^{\dot n_k-\dot \ell_k} \le(\frac {\ds
  \Delta_{n_k}^{\ell_k}} {\ds
  \Delta_{n_{_{k-1}}}^{\ell_{_{k-1}}}}\ri)^2\\
&&\prod_{j=1}^{k}  \det T_j(x) =  (-1)^{n_k} x^{n_k-\ell_k}  \le( \Delta_{n_k}^{\ell_k}\ri)^2
\eea
Summing up both sides of eq. (\ref{ads}) we obtain the following {\em master Christoffel--Darboux identity}
\bea
&&\Xi_{N}^\star(x)\,\mathbf j \,\Xi_{N}(y) - \le[\matrix{0 & -1/y\cr 1/x & 0} \ri] =\nonumber\\
&&=\le(\frac 1 y -\frac 1
x\ri)\sum_{k=0}^{N-1}(-1)^{1-\dot n_{k+1}} x^{\dot \ell_{k+1}}  y^{\dot n_{k+1}}
 \Xi_{k}^\star(x)\,
\le[\begin{array}{cc}
\ds \frac{\Delta_{n_{k+1}-1}^{\ell_{k+1}-1}}{\Delta_{n_{k+1}}^{\ell_{k+1}}} &
1-\dot n_{k+1} \\[15pt]
 \dot \ell_{k+1}-1 & 0
\end{array}\ri]
\,\Xi_{k}(y)=\nonumber\\
&&=\le(\frac 1 x -\frac 1
y\ri)\sum_{k=0}^{N-1}(-1)^{-\dot n_{k+1}} x^{\dot \ell_{k+1}}  y^{\dot n_{k+1}}
 \Xi_{k}^\star(x)\,
\le[\begin{array}{cc}
\ds \frac{\Delta_{n_{k+1}-1}^{\ell_{k+1}-1}}{\Delta_{n_{k+1}}^{\ell_{k+1}}} &
1-\dot n_{k+1} \\[15pt]
 \dot \ell_{k+1}-1 & 0
\end{array}\ri]
\,\Xi_{k}(y)\label{master CDI}
\eea
\subsection{Principal CDI}
We look at the $(1,1)$ entry of the above identity
\bea
&&\frac {(-1)^{n_N}}{(\Delta_{n_N}^{\ell_N})^2} \le(
\frac{\wp_{n_N-1}^{\ell_N-1}(x)}{x^{n_N-\ell_N}}\wp_{n_N}^{\ell_N}(y)
- \frac{\wp_{n_N}^{\ell_N}(x)}{x^{n_N-\ell_N}}\wp_{n_N-1}^{\ell_N-1}(y)
\ri)= \le(\frac 1 x -\frac 1 y\ri)
\sum_{k=0}^{N-1} \frac {(-1)^{n_{k+1}} x^{\ell_{k+1}-n_{k}}y^{\dot n_{k+1}}}
    {(\Delta_{n_{k}}^{\ell_{k}})^2}\times \nonumber \\
&&\times \le[
\frac{\Delta_{n_{k+1}-1}^{\ell_{k+1}-1}}{\Delta_{n_{k+1}}^{\ell_{k+1}}}
\wp_{n_k}^{\ell_k}(y) + (1-\dot n_{k+1})\wp_{n_k-1}^{\ell_k-1}(y)\ri] \le[
\wp_{n_k}^{\ell_k}(x) - (1-\dot \ell_{k+1})
\frac{\Delta_{n_{k+1}}^{\ell_{k+1}}}{\Delta_{n_{k+1}-1}^{\ell_{k+1}-1}}
  \wp_{n_k-1}^{\ell_k-1}(x)
\ri]\label{pcdi}
\eea
The two terms in
the product inside the sum here above can be simplified  using
(\ref{uno}) for the case $\dot \ell_{k+1}=0$ and (\ref{tre}) for the
case $\dot n_{k+1}=0$
indeed
\bea
\le[
  \frac{\Delta_{n_{k+1}-1}^{\ell_{k+1}-1}}{\Delta_{n_{k+1}}^{\ell_{k+1}}}
\wp_{n_k}^{\ell_k}(y) + (1-\dot n_{k+1})\wp_{n_k-1}^{\ell_k-1}(y)\ri]
&=& \le\{
\begin{array}{ll}
\ds y\frac {\Delta^{\ell_k}_{n_k}}{\Delta_{n_k}^{\ell_k+1}}
\wp_{n_{k+1}-1}^{\ell_{k}}(y) & \hbox{ if } \dot n_{k+1}=0\\[20pt]
\ds\frac{\Delta_{n_{k+1}-1}^{\ell_{k+1}-1}}{\Delta_{n_{k+1}}^{\ell_{k+1}}}
\wp_{n_k}^{\ell_k}(y)  & \hbox{ if } \dot n_{k+1}=1
\end{array}\ri.\nonumber\\
&=& y^{1-\dot n_{k+1}} \frac
       {\Delta_{n_k}^{\ell_{k+1}-1}}{\Delta_{n_{k+1}}^{\ell_{k+1}}
       } \wp_{n_{k+1}-1}^{\ell_{k}}(y)
\eea
\bea
\le[
\wp_{n_k}^{\ell_k}(x) - (1-\dot \ell_{k+1})
\frac{\Delta_{n_{k+1}}^{\ell_{k+1}}}{\Delta_{n_{k+1}-1}^{\ell_{k+1}-1}}
  \wp_{n_k-1}^{\ell_k-1}(x)
\ri]
&=&
 \le\{
\begin{array}{ll}
\ds \frac {\Delta_{n_k}^{\ell_{k+1}}}{\Delta_{n_{k+1}-1}^{\ell_{k+1}-1}}
\wp_{n_k}^{\ell_{k+1}-1}(x) &\hbox { if } \dot \ell_{k+1}=0 \\[20pt]
 \ds\wp_{n_k}^{\ell_k}(x) =\wp_{n_k}^{\ell_{k+1}-1}(x) & \hbox{ if } \dot \ell_{k+1}=1
\end{array}\ri.\nonumber\\
&=& \frac {\Delta_{n_k}^{\ell_k}}{\Delta_{n_k}^{\ell_{k+1}-1}} \wp_{n_k}^{\ell_{k+1}-1}(x)
\eea
Using these expression in the RHS of (\ref{pcdi}) the identity becomes
\bea
\frac {(-1)^{n_N}}{(\Delta_{n_N}^{\ell_N})^2} \le(
\frac{\wp_{n_N-1}^{\ell_N-1}(x)}{x^{n_N-\ell_N}}\wp_{n_N}^{\ell_N}(y)
- \frac{\wp_{n_N}^{\ell_N}(x)}{x^{n_N-\ell_N}}\wp_{n_N-1}^{\ell_N-1}(y)
\ri)
= \nonumber \\ =
\le( \frac y x -1 \ri)\sum_{k=0}^{N-1} {(-1)^{n_{k+1}}}
\frac{\wp_{n_{k+1}-1}^{\ell_k}(y)
  \wp_{n_k}^{\ell_{k+1}-1}(x)x^{\ell_{k+1}-n_{k}}}     {\Delta_{n_{k+1}}^{\ell_{k+1}} \Delta_{n_k}^{\ell_k}}
\eea
We can repeat the same arguments for the second-kind polynomials appearing
in the other matrix entries; care must be paid to the fact that
$(\Xi_0)_{12}$ is not $\wr_0^0\equiv 0 $.

 We obtain the following supplementary
CDI's (provided that $0\leq \ell_k \leq n_{k+1}-2$, $k=1,\dots$)
\bea
&&\hspace{-1cm}
\frac {(-1)^{n_N}}{(\Delta_{n_N}^{\ell_N})^2} \le(
\frac{\wr_{n_N-1}^{\ell_N-1}(x)}{x^{n_N-\ell_N}}\wp_{n_N}^{\ell_N}(y)
- \frac{\wr_{n_N}^{\ell_N}(x)}{x^{n_N-\ell_N}}\wp_{n_N-1}^{\ell_N-1}(y)
\ri) -\frac 1 x
=\nonumber \\
&&=
\le( \frac y x -1 \ri)\sum_{k=0}^{N-1} {(-1)^{n_{k+1}}}
\frac{\wp_{n_{k+1}-1}^{\ell_k}(y)
  \wr_{n_k}^{\ell_{k+1}-1}(x)x^{\ell_{k+1}-n_{k}}}
     {\Delta_{n_{k+1}}^{\ell_{k+1}} \Delta_{n_k}^{\ell_k}} \nonumber \\
&&\hspace{-1cm}
\frac {(-1)^{n_N}}{(\Delta_{n_N}^{\ell_N})^2} \le(
\frac{\wp_{n_N-1}^{\ell_N-1}(x)}{x^{n_N-\ell_N}}\wr_{n_N}^{\ell_N}(y)
- \frac{\wp_{n_N}^{\ell_N}(x)}{x^{n_N-\ell_N}}\wr_{n_N-1}^{\ell_N-1}(y)
\ri)+\frac 1 y
=\nonumber \\
&& =
\le( \frac y x -1 \ri)\le[\sum_{k=0}^{N-1} {(-1)^{n_{k+1}}}
\frac{\wr_{n_{k+1}-1}^{\ell_k}(y)
  \wp_{n_k}^{\ell_{k+1}-1}(x)x^{\ell_{k+1}-n_{k}}}
     {\Delta_{n_{k+1}}^{\ell_{k+1}} \Delta_{n_k}^{\ell_k}} - \frac 1
     y\ri]\nonumber \\
&&\hspace{-1cm}
\frac {(-1)^{n_N}}{(\Delta_{n_N}^{\ell_N})^2} \le(
\frac{\wr_{n_N-1}^{\ell_N-1}(x)}{x^{n_N-\ell_N}}\wr_{n_N}^{\ell_N}(y)
- \frac{\wr_{n_N}^{\ell_N}(x)}{x^{n_N-\ell_N}}\wr_{n_N-1}^{\ell_N-1}(y)
\ri)
=\nonumber\\
&&=
\le( \frac y x -1 \ri) \sum_{k=0}^{N-1} {(-1)^{n_{k+1}}}
\frac{\wr_{n_{k+1}-1}^{\ell_k}(y)
  \wr_{n_k}^{\ell_{k+1}-1}(x)x^{\ell_{k+1}-n_{k}}}     {\Delta_{n_{k+1}}^{\ell_{k+1}} \Delta_{n_k}^{\ell_k}}
\eea
The additional term in the second identity stems from the mentioned
discrepancy in the definition of $\Xi_0$ with the definition of the
auxiliary polynomials: indeed the term with $k=0$ in the sum
(\ref{master CDI}) is not zero in the off-diagonal terms but
$\le[\matrix{1 & -\frac 1 y \cr 0 & 0}\ri]$.
Thus the second identity above is rewritten as
\bea
&&\frac {(-1)^{n_N}}{(\Delta_{n_N}^{\ell_N})^2} \le(
\frac{\wp_{n_N-1}^{\ell_N-1}(x)}{x^{n_N-\ell_N}}\wr_{n_N}^{\ell_N}(y)
- \frac{\wp_{n_N}^{\ell_N}(x)}{x^{n_N-\ell_N}}\wr_{n_N-1}^{\ell_N-1}(y)
\ri)+\frac 1 x
=\nonumber \\&&\hspace{2cm} =
\le( \frac y x -1 \ri)\sum_{k=0}^{N-1} {(-1)^{n_{k+1}}}
\frac{\wr_{n_{k+1}-1}^{\ell_k}(y)
  \wp_{n_k}^{\ell_{k+1}-1}(x)x^{\ell_{k+1}-n_{k}}}
     {\Delta_{n_{k+1}}^{\ell_{k+1}} \Delta_{n_k}^{\ell_k}}
\eea
\section{CDIs for biorthogonal Laurent polynomials}
The formul\ae\ derived in the previous sections for the
Christoffel-Darboux identities are very general however the (Laurent)
polynomials that appear
in the sum are not biorthogonal  with respect to the  moment functional $\mathcal L$ unless the sequence $n_k$
is strictly increasing and the sequence $\ell_k$ is weakly increasing. This is the situation which interests us the
most and hence from now on we will assume that $n_k =k$\footnote{If
  $n_k$ were not strictly increasing then the
polynomials would be biorthogonal only provided the moments satisfy some
non generic condition of vanishing of certain determinants.}. Moreover
all the elementary orbits of the integrable lattices we are
considering are in correspondence with this situation.

From the formul\ae\ defining the polynomials $\wp_n^\ell$ it follows that
\be
\mathcal L_z \le(\wp_{n}^{\ell_n}(z) \wp_{m}^{\ell_{m+1}-1}(z)
z^{\ell_{m+1}-m}\ri) = \delta_{mn} (-1)^n \Delta_{n}^{\ell_n}
\Delta_{n+1}^{\ell_{n+1}}\ .
\ee
This suggests that we introduce the following monic polynomials
\bea
&&\pi_n(x) = \frac{1}{\Delta_n^{\ell_n}} \wp_n^{\ell_n}(x)\\
&&\rho_n(x)= \frac{
  (-1)^n}{\Delta_n^{\ell_n}}x^{\ell_{n+1}-n}\wp_n^{\ell_{n+1}-1}(x)\ .
\eea
It is understood that the determinants $\Delta_n^{\ell_n}$ must not vanish: this is our implicit assumption of genericity on the moment functional. 
While the $\pi_n$'s are monic in the usual sense, the $\rho_n$'s are
normalized on either the highest or the lowest power depending on
$\dot \ell_{n+1}$. Moreover  the $\pi_n$'s are polynomials in $x$
whereas the $\rho_n$'s are polynomials in $x$ and $x^{-1}$.
They satisfy the orthogonality relations
\be
\mathcal L_z(\rho_m(z)\pi_n(z)) = \delta_{mn} h_n\ ,\qquad h_n:= \frac
     {\Delta_{n+1}^{\ell_{n+1}}}{\Delta_n^{\ell_n}}\ .
\ee
We finally introduce the {\bf  (bi)-orthonormal polynomials} and the second kind polynomials
\bea
&& p_n(x) := \frac  1 {\sqrt{h_n}} \pi_n(x) = \frac
    {\wp_n^{\ell_n}}{\sqrt{\Delta_{n}^{\ell_n}
    \Delta_{n+1}^{\ell_{n+1}}}}  ,\nonumber\\
&&\wt p_n(x):= \mathcal L_z\le(\frac {p_n(x)-p_n(z)}{x-z} \ri)\nonumber\\
&& r_n(x) := \frac  1 {\sqrt{h_n}} \rho_n(x) =x^{\ell_{n+1}-n}  \frac{(-1)^n
      \wp_{n}^{\ell_{n+1}-1}} {\sqrt{\Delta_{n}^{\ell_n}
    \Delta_{n+1}^{\ell_{n+1}}}}\nonumber \\
&&\wt r_n(x):= \mathcal L_z\le(\frac {r_n(x)-r_n(z)}{x-z} \ri)\label{pirho}
\eea
and their ``starred''
\bea
p_n^\star(x) :=x^{\ell_n -n+1} p_n(x)\qquad \wt p_n^\star(x) :=x^{\ell_n -n+1}  \wt p_n(x)\\
r_n^\star(x) := x^{n-\ell_{n+1}}r_n(x)\qquad \wt r_n^\star(x) := x^{n-\ell_{n+1}} \wt r_n(x)
\eea
In terms of these (Laurent)polynomials the CDIs read
\bea
\le(y-x\ri) \sum_{n=0}^{N-1} r_n(x) p_n(y)& =&
\gamma_N \le( p_N(y)r_{N-1}(x) - p_N^\star(x)
r_{N-1}^\star(y)\ri) \nonumber\\
\le(y-x\ri) \sum_{n=0}^{N-1}\wt r_n(x) p_n(y) &=&
\gamma_N  \le( p_N(y)\wt r_{N-1}(x) - \wt p_N^\star(x)
r_{N-1}^\star(y)\ri) +1 \nonumber\\
\le(y-x\ri) \sum_{n=0}^{N-1} r_n(x)\wt  p_n(y) &=&
\gamma_N   \le( \wt p_N(y)r_{N-1}(x) - p_N^\star(x)
\wt r_{N-1}^\star(y)\ri) - 1 \nonumber\\
\le(y-x\ri) \sum_{n=0}^{N-1}\wt  r_n(x)\wt  p_n(y) &=&
\gamma_N \le(\wt  p_N(y)\wt r_{N-1}(x) - \wt p_N^\star(x)
\wt r_{N-1}^\star(y)\ri)\nonumber\\
&& \gamma_N:= \sqrt{\frac{h_{N}}{h_{N-1}}} \label{CDIs}
\eea
It is convenient to rewrite in matrix form the previous identities as
follows
\bea
&& \p (x) :=[p_0,\dots]^t\ ,\ \wt \p(x)  := [\wt p_0,\dots]^t\ ,\ \r(x) :=
[r_0,\cdots]^t\ ,\ \wt\r(x) :=
[\wt r_0,\cdots]^t\\
&&{\bf P}(x):= \le[\p(x),\wt\p(x)\ri]\ ,\qquad {\bf R}(x):=
\le[\r(x),\wt \r(x)\ri]\\
&&(\Pi_{N-1})_{ij}:= \sum_{k=0}^{N-1} \delta_{ik}\delta_{kj}\\
&&{\bf R}^t(x) \Pi_{N-1} {\bf P}(y) =\frac 1{y-x}\le\{\gamma_N \le[
\matrix{p_N^\star(x) & r_{N\!-\!1}(x)\cr
{\wt p}^\star _N(x) & \wt r_{N\!-\!1}(x)}
\ri]{\mathbf j}\le[\matrix{p_N(y) & \wt p_{N}(y)\cr
{r}^\star_{N\!-\!1} (y) & {\wt r}^\star_{N\!-\!1}(y)}
\ri] + \mathbf j\ri\}
\eea
\br
A word about the relations with previously known (bi)-orthogonal
polynomials is now in order. If all the moves (except the first one)
are {\em line-moves} namely if $\ell_n =  n-1$ then it is not hard to
show that $\pi_n = \rho_n$ are just orthogonal polynomials with
respect to the (restriction of the) moment functional $\mathcal L$ to
positive moments. Moreover the shifted T\"oplitz determinants $\Delta_n^{n-1}$ are  (up to a sign) the same as the H\"ankel determinants of the same size (by permuting appropriately the columns).

Vice-versa, if all moves are {\em circle-moves}
(i.e. $\ell_n \equiv 0$) (and we also impose certain reality
conditions on the moments of the functional) then the $\pi_n$ are
orthogonal polynomials for a certain measure on the unit circle and
the $\rho_n$'s are their so-called ``dual'' Laurent polynomials. The determinants appearing then in our sequence are precisely the ``standar'' ones $\Delta_n^{0}$.

A second remark is that all these polynomial {\em do} satisfy three-terms recurrence relations, although of a different sort  than the standard ones.
Indeed, it is well known that orthogonal polynomials $p_n$ satisfy relations of the form 
\be
xp_n = \gamma_n p_{n+1}+ \beta_n p_n + \gamma_{n-1} p_{n-1}\ ,
\ee
where the coefficients $\gamma_n,\beta_n$ enter in the tridiagonal Jacobi matrix representing the multiplication by $x$ in the basis of the $p_n$'s. At the opposite "end of the spectrum", orthogonal polynomials on the circle satisfy a different sort of three term recurrence relation, of the form
\be
x(p_n + \delta_n p_{n-1}) = \gamma_n p_{n+1}  + \beta_n p_n\ .
\ee
It is not hard to show \cite{FG1,FG2} that the polynomials that we are considering precisely "interpolate" these two sorts of recurrence relations as follows
\be
x(p_n + (1-\dot \ell_n) \delta_n p_{n-1}) = \gamma_n p_{n+1} + \dot \ell_n \beta_np_n\ ,
\ee
for certain coefficients $\gamma_n,\beta_n,\delta_n$ whose explicit expression in terms of T\"oplitz determinants can be obtained from the formul\ae\ above but is irrelevant for this discussion. We see that ``circle moves'' ($\dot \ell_n=0$) correspond to a three-term recurrence relation of the type appearing for O.P. on the circle, while ``line moves'' ($\dot \ell_n=1$) correspond to the ``usual'' recurrence relation. 
\er 
\section{Infinitesimal deformations of the moment functional}
We study the  infinitesimal deformations for the wave vectors $\p (x)
$, $\wt \p(x)$, $\r(x)$ and $\wt\r(x)$ under an infinitesimal
deformation of the moment functional. Let us introduce the matrix of
recurrence for these sequences of polynomials
\bea
x \p = Q \p\ ;\ x \r^t= \r^t Q\ ,\qquad Q_{nm}:= \mathcal L (z\,p_n\,
r_m)\ .
\eea
The matrix $Q$ is of Hessenberg form, namely has
nonzero entries on the superdiagonal and possibly on the diagonal and
all other nonzero entries in the lower triangular part.
The biorthogonality relation can be rewritten as
\bea
\mathcal L\le[ \p \r^t\ri] = \1
\eea
Suppose we infinitesimally deform the moment functional
\be
\dot{\mathcal L}(\bullet) = - \mathcal
L( F(z)\bullet)
\ee
Here $F(z)$ can be any function (even a generalized distribution as we will see) provided that the moments of the
deformation are still well defined: if $\mathcal L$ is given by an
analytical expression in terms of some integral representation (as we will
assume later on) then this means some condition of analyticity on $F$:
 if the functional is only defined by its moments, then $F$
should be interpreted as formal series. In any situation  the typical case
of $F$ being a polynomial (corresponding to the usual formal Toda-type flows) will be well defined.

A little more generally we could even assume that $F$ is a
distribution, particularly delta functions or derivatives of it.
For instance we can consider deformation of the type
\be
\delta\mathcal L(p(x))\equiv \dot{\mathcal L}(p(x)) = -\le(\frac {\rm
  d}{{\rm d}x}\ri)^k p(x)\Bigg|_{x=a}
\ee
for some constant  $a$: this means that we (formally) have set $F$ to
be the $k$-th derivative of the Dirac delta distribution for the given
moment functional supported at $x=a$.\\
Corresponding to any of these  deformations  the BOPs deform as
\be
\delta \p = \mathbb U^{(F)} \p\ ,\ \ \delta \r = \wt{\mathbb U}^{(F)}\r
\ee
where a priori $\mathbb U$ and $\wt {\mathbb U}$ are lower triangular
matrices since the range of powers of $x$ entering in the expressions
$p_n$, $r_n$ will not change.
In order to find expressions for these matrices we note first that  their diagonals are the same
\be
(\mathbb U^{(F)})_{nn}= (\wt{\mathbb U}^{(F)})_{nn} = -\frac 1 2
\delta{\ln(h_n)} \label{hns}
\ee
Indeed we have
\bea
\delta p_n = \delta \frac {x^n}{\sqrt{h_n}} + \dots=  -\frac 1 2
\delta \ln(h_n) p_n + \hbox{ previous }\\
\delta r_n = -\frac 1 2
\delta \ln(h_n) r_n + \hbox{ previous }\ .
\eea
Differentiating the orthogonality relation  we obtain
\be
\mathbb U^{(F)}+ {\wt{\mathbb U}^{(F)}}\!^t = \le\{\begin{array}{cc}
\ds F(Q) & \hbox{ for the case of an ordinary function $F$}\\[20pt]
\ds\le(\frac {\rm
  d}{{\rm d}x}\ri)^k \p(x)\r^t(x)\Bigg|_{x=a} & \hbox{ for a
  deformation supported at one point}
\end{array}\ri.
\ee
and hence according to the two types the matrices describing the
infinitesimal deformations are given by
\bea
 &\mathbb U^{(F)} = F(Q)_{-0}\ ,\qquad &\wt{\mathbb U}^{(F)} = {F(Q)^t}_{-0}
\label{deform}\\
& \mathbb U^{(\delta_a^k)} =\pa_a^k\le(\p(a)\r^t(a)\ri)_{-0} \
,\qquad
&\wt{\mathbb U}^{(\delta_a^k)} =\pa_a^k \le(\r (a)\p^t(a)\ri)_{-0}
\eea
where $A_{-0}$ means the lower triangular part plus half of the
diagonal.
Note that from (\ref{hns}) and the definition of $h_n$ it follows
\bea
&& \Delta_n^{\ell_n} = \prod_{k=0}^{n-1} h_k\nonumber \\
&& \delta_f \ln \Delta_n^{\ell_{n}} = -\tr_n F(Q)\nonumber \\
&& \delta_{\delta_a^k} \ln \Delta_n^{\ell_{n}} =
-\pa_a^k\sum_{j=0}^{n-1} p_j(a)r_j(a)\label{defDelta}\ ,
\eea
where we have used the notation for the {\bf truncated trace} $\tr_n A:=\sum_{j=0}^{n-1} A_{jj}$.
\subsection{Deformations for the second-kind (Laurent) polynomials}
Using Leibnitz's rule we obtain the following deformation
equations for the second-kind wave vectors $\wt \p$, $\wt\r$.
For a deformation by a function $F(x)$ we have
\bea
\delta_F \wt \p &=& \le(\mathbb U^{(F)} -F(x)\ri) \wt \p + \mathcal
L_z\le(\frac {F(x)-F(z)}{x-z} \ri)\p - \le(\frac
{F(x)-F(Q)}{x-Q}\ri)\mathbf e_1 \nonumber \\
\delta_F \wt \r &=& \le(\wt{\mathbb U}^{(F)} -F(x)\ri) \wt \r + \mathcal
L_z\le(\frac {F(x)-F(z)}{x-z} \ri)\r - \le(\frac
{F(x)-F(Q^t)}{x-Q^t}\ri)\mathbf e_1 \label{def2nd1}
\eea
while for $F= \delta^{(k)}_{\mathcal L}(z-a)$ we have
\bea
\delta_F \wt \p &=& \mathbb U^{(\delta_a^{k})} \wt p -
\frac{{\pa}^k}{{\pa }a^k} \frac {\p(x)-\p(a)}{x-a}\nonumber \\
\delta_F \wt \r &=& \wt{\mathbb U}^{(\delta_a^{k})} \wt r - \frac
      {\pa^k}{\pa a^k} \frac {\r(x)-\r(a)}{x-a}\label{def2nd2}
\eea
\section{Folded version of the deformation equations}
Let us define
\be
\chi_n:=\le[\matrix{ p_n & \wt p_n\cr r_{n-1}^\star  & {\wt r}_{n-1}^\star}\ri]
\ee
We want to express the previous infinite-dimensional deformation
equations in terms of $\chi_n$ alone; this process is conceptually
identical to the one followed in \cite{semiiso} and which is named
"folding". To this end we formulate the
following
\bt
The infinite deformations (\ref{deform}) for the wave vectors $\p,\r$
and for the second-kind wave vectors $\wt \p, \wt\r$ (\ref{def2nd1},
\ref{def2nd2}) are equivalent to the following deformation equations
for $\chi_n$, $n\geq 1$.
\bea
\delta_{(F)}\chi_n &=& \mathcal U^{(F)}_n(x) \chi_n + \chi_n  {\mathcal
  U}^{(F),R}(x)\nonumber\\[10pt]
  \delta_{(\delta_a^k)} \chi_n(a) &=& \mathcal U^{(\delta_a^k)}_n(x)
\chi_n(x) + \chi_n(x)\mathcal  U^{(\delta_a^k),R}(x)
 \eea
 where we have used the following definitions:
\bea
\mathcal U^{(F)}_n &=&  \le[\begin{array}{cc}
\frac 1 2 F(Q)_{nn} & 0 \\
0& F(x)-\frac 1 2 F(Q)_{n\!-\!1,n\!-\!1}
\end{array}
\ri] + \gamma_n \le[
\begin{array}{cc}
-(\nabla_QF)_{n,n\!-\!1} & (\nabla_Q F)_{n,n^\star}\\
-(\nabla_QF)_{(n\!-\!1)^\star,n\!-\!1} & (\nabla_QF)_{n,n\!-\!1}
\end{array}
\ri]
\nonumber \\[10pt]
\mathcal U^{(F),R} &=& \le[\begin{array}{cc}
0& \mathcal W_F\\
0& -F(x)
\end{array}\ri]\ , \ \mathcal W_F:= \mathcal L_z\le( \frac
     {F(x)-F(z)}{x-z}\ri)\ ,\qquad \nabla_Q F := \frac {F(x)-F(Q)}{x-Q}
\\[20pt]
\mathcal  U^{(\delta_a^k)}_n(x) &=& \frac {\pa^k}{\pa a^k}  \frac 1 2\le[\begin{array}{cc}
p_nr_n & 0\\
0& -p_{n\!-\!1}r_{n\!-\!1}
\end{array}\ri]_{z=a} + \frac {\pa^k}{\pa a^k}\frac {\gamma_n}{x-a} \le[\begin{array}{cc}
- p_n r_{n\!-\!1} & p_n p_n^\star \\
-  r_{n\!-\!1}  r_{n\!-\!1}^\star & r_{n\!-\!1} p_n
\end{array}\ri]_{z=a}
\nonumber \\[10pt]
\mathcal  U^{(\delta_a^k),R}(x) &=& \pa_a^k\le[\matrix{0 & \frac
    1{a-x}\cr 0&0}\ri]\ .
\eea
Here, for a function $f(z)$ we have set
\be
f(Q)_{i,j^\star} := \mathcal L(r_i f(z) r_i^\star)\ ,\ \
f(Q)_{i^\star,j}:= \mathcal L (p_i^\star f(z)p_j)\ .
\ee
\et
{\bf Proof.}
 We compute the
deformations of both rows of $\chi_n$.
We start with  deformation involving a function $F(x)$:
 the first row deforms according
to the equation
\bea
\delta_F \le[p_n(x),\wt p_n(x)\ri] = \delta_F \e_n^t\cdot [\p,\wt \p] =
\e_n^t \cdot\mathbb U^{(F)} \cdot  [ \p,\wt\p]\  + \   \e_n^t\cdot
[\p.\wt \p]\le[\matrix{0& \mathcal W_F\cr0&  -F(x)} \ri]
 \ -\  \e_n^t\cdot\frac{F(x)- F(Q)}{x-Q}\cdot[\mathbf 0,\e_1]\ ,
\eea
where we have set
\be
\mathcal W_F(x):= \mathcal L_z\le(\frac {F(x)-F(z)}{x-z}\ri)
\ee
We now note that
\bea
\e_n^t \cdot\mathbb U^{(F)} \cdot  [ \p,\wt\p] = \frac 1 2 F(Q)_{nn}[p_n,\wt
  p_n] +\e_n^t \mathcal L_z\le(F(z) \p(z) \r^t(z) \Pi_{n-1}   [
  \p(x),\wt\p(x)]\ri) = \nonumber\\
=  \frac 1 2 F(Q)_{nn}[p_n,\wt
  p_n] +\e_n^t \mathcal L_z\le((F(z)-F(x)) \p(z) \r^t(z) \Pi_{n-1}   [
  \p(x),\wt\p(x)]\ri)=\nonumber\\
= \frac 1 2 F(Q)_{nn}[p_n,\wt
  p_n] +\e_n^t \mathcal L_z\le(\frac{F(z)-F(x)}{x-z} \p(z)
\bigg(\gamma_n [p_n(z)^\star,r_{n-1}(z)] \mathbf j \chi_n(x) -
     [0,1]\bigg)\ri) =\nonumber\\
= \frac 1 2 F(Q)_{nn}[p_n,\wt
  p_n] - \gamma_n \mathcal L_z\le(\frac{F(z)-F(x)}{z-x}
p_n [p_n^\star,r_{n-1}] \ri) \mathbf j \chi_n(x) +
\e_n^t\cdot\frac{F(x)- F(Q)}{x-Q}\cdot[\mathbf 0,\e_1]
\eea
This implies that
\bea
\delta_F \le[p_n(x),\wt p_n(x)\ri] = \frac 1 2 F(Q)_{nn}[p_n,\wt
  p_n] - \gamma_n \mathcal L_z\le(\frac{F(z)-F(x)}{z-x}
p_n[p_n^\star,r_{n-1}] \ri) \mathbf j \chi_n(x) +
\nonumber\\+ \   \e_n^t\cdot
[\p.\wt \p]\le[\matrix{0& \mathcal W_F\cr0&  -F(x)} \ri]\label{def1}
\eea
In a similar way we can compute the following deformations
\bea
&&\delta_F \le[r_{n\!-\!1}(x),\wt r_{n\!-\!1} (x)\ri] =
\delta_F \e_{n\!-\!1}^t\cdot [\r,\wt \r] =\nonumber \\&&= 
\e_{n\!-\!1}^t \cdot {\mathbb U^{(F)}}\!^t\cdot  [ \r,\wt\r]\  + \   \e_{n\!-\!1}^t\cdot
[\r,\wt \r]\le[\matrix{0& \mathcal W_F\cr0&  -F(x)} \ri]
 \ -\  \e_{n\!-\!1}^t\cdot\frac{F(x)- F(Q^t)}{x-Q^t}\cdot[\mathbf 0,\e_1]\ .
\eea
The computation now involves
\bea
\e_{n\!-\!1}^t \cdot {\mathbb U^{(F)}}\!^t\cdot  [ \r,\wt\r]& =& -\frac 1 2
F(Q)_{n\!-\!1,n\!-\!1}  [ r_{n\!-\!1},\wt r_{n\!-\!1}]\  +\
\e_{n\!-\!1}^t \cdot F(Q^t) \Pi_{n\!-\!1} [ \r,\wt\r]=\nonumber\\
&=&-\frac 1 2
F(Q)_{n\!-\!1,n\!-\!1}  [ r_{n\!-\!1},\wt r_{n\!-\!1}]\  +\
\e_{n\!-\!1}^t \mathcal L_z\le( F(z)\r(z)\p^t(z) \Pi_{n\!-\!1} [
  \r(x),\wt\r(x)]\ri)=\nonumber\\
&=&\le(F(x)-\frac 1 2
F(Q)_{n\!-\!1,n\!-\!1}\ri)  [ r_{n\!-\!1},\wt r_{n\!-\!1}]\
+\nonumber \\
&&+ \,\e_{n\!-\!1}^t \mathcal L_z\le( (F(z)-F(x)) \r(z)\p^t(z) \Pi_{n\!-\!1} [
  \r(x),\wt\r(x)]\ri) = \nonumber\\[10pt]
&=&\le(F(x)-\frac 1 2
F(Q)_{n\!-\!1,n\!-\!1}\ri)  [ r_{n\!-\!1},\wt r_{n\!-\!1}]\  +\nonumber\\
&& + \e_{n\!-\!1}\mathcal L_z\le( \frac{F(z)-F(x)}{z-x} \r(z)
\bigg(-\gamma_n[p_n(z), r_{n\!-\!1}^\star(z)] \mathbf j {\chi_n^\star(x)}^t
+[0,1]\bigg)\ri)=\nonumber\\
&=& \le(F(x)-\frac 1 2
F(Q)_{n\!-\!1,n\!-\!1}\ri)  [ r_{n\!-\!1},\wt r_{n\!-\!1}]
+\nonumber\\
&& -\gamma_n  \mathcal L_z\le( \frac{F(z)-F(x)}{z-x} r_{n\!-\!1} [p_n
  , r_{n\!-\!1}^\star ]\ri) \mathbf j {\chi_n^\star}^t(x) +
\e_{n\!-\!1}\frac{F(x)-F(Q^t)}{x-Q^t}\e_1
\eea
where we have used the following definition
\be
\chi_n^\star = x^{\ell_n-n+1}\chi^t_n\ .
\ee
Thus we have obtained  the following deformation equation
\bea
\delta_F \le[r_{n\!-\!1}(x),\wt r_{n\!-\!1} (x)\ri] &=& \le(F(x)-\frac 1 2
F(Q)_{n\!-\!1,n\!-\!1}\ri)  [ r_{n\!-\!1},\wt r_{n\!-\!1}]   \ +\  \e_{n\!-\!1}^t\cdot
[\r,\wt \r]\le[\matrix{0& \mathcal W_F\cr0&  -F(x)} \ri]+\nonumber \\&&
 -\gamma_n  \mathcal L_z\le( \frac{F(z)-F(x)}{z-x} r_{n\!-\!1} [p_n
  , r_{n\!-\!1}^\star ]\ri) \mathbf j {\chi_n^\star}^t(x)
\eea
By ``starifying'' both sides we obtain
\bea
\delta_F \le[r_{n\!-\!1}^\star(x),\wt r_{n\!-\!1}^\star (x)\ri] &=& \le(F(x)-\frac 1 2
F(Q)_{n\!-\!1,n\!-\!1}\ri)  [ r_{n\!-\!1}^\star,\wt r_{n\!-\!1}^\star]   \ +\
[r_{n\!-\!1}^\star,\wt r_{n\!-\!1}^\star]\le[\matrix{0& \mathcal W_F\cr0&  -F(x)} \ri]+\nonumber \\&&
 -\gamma_n  \mathcal L_z\le( \frac{F(z)-F(x)}{z-x} r_{n\!-\!1} [p_n
  , r_{n\!-\!1}^\star ]\ri) \mathbf j \chi_n(x) \label{def2}
\eea
Putting together (\ref{def1}) and (\ref{def2}) we obtain finally
\bea
\delta_F\chi_n &=& \mathcal U^{(F)}_n(x) \chi_n + \chi_n  {\mathcal
  U}^{(F),R}(x)\\
\mathcal U^{(F)}_n &=& \le[\begin{array}{cc}
\frac 1 2 F(Q)_{nn} & 0 \\
0& F(x)-\frac 1 2 F(Q)_{n\!-\!1,n\!-\!1}
\end{array}
\ri]
 -\gamma_n \mathcal L_z\le(\frac{F(x)-F(z)}{x-z} \le[
\begin{array}{cc}
p_np_n^\star & p_nr_{n\!-\!1}\\
p_n r_{n\!-\!1} & r_{n\!-\!1}^\star  r_{n\!-\!1}
\end{array}\ri]\ri)\mathbf j\nonumber =\\
&=& \le[\begin{array}{cc}
\frac 1 2 F(Q)_{nn} & 0 \\
0& F(x)-\frac 1 2 F(Q)_{n\!-\!1,n\!-\!1}
\end{array}
\ri] + \gamma_n \le[
\begin{array}{cc}
-(\nabla_QF)_{n,n\!-\!1} & (\nabla_Q F)_{n,n^\star}\\
-(\nabla_QF)_{(n\!-\!1)^\star,n\!-\!1} & (\nabla_QF)_{n,n\!-\!1}
\end{array}
\ri]\\
\mathcal U^{(F),R} &=& \le[\begin{array}{cc}
0& \mathcal W_F\\
0& -F(x)
\end{array}\ri]
\eea
We now consider a deformation supported at one point $z=a$.
\bea
\delta_F \le[p_n(x),\wt p_n(x)\ri] =
\e_n^t \cdot\mathbb U^{(\delta_a^k)} \cdot  [ \p,\wt\p]\  -\    \e_n^t\le(\frac{{\rm d}}{{\rm
    d}z}\ri)^k\bigg|_{z=a} \frac {\p(x)-\p(z)}{x-z}[0,1]\ .
\eea
This time we have
\bea
\e_n^t \cdot\mathbb U^{(\delta_a^k)} \cdot  [ \p,\wt\p] = \frac 12 \pa_a^k(p_n(a)r_n(a))
[p_n,\wt p_n] + \pa_a^k \e_n^t \cdot \p(a) \r^t(a)\Pi_{n\!-\!1}
[\p,\wt\p] =\\
=\frac 12 \pa_a^k(p_n(a)r_n(a))
[p_n,\wt p_n] +\pa_a^k \e_n^t \cdot \p(a)\le(\frac
{\gamma_n[p_n^\star(a),r_{n\!-\!1}(a)]}{x-a} \mathbf j\,\chi_n(x) -\frac {[0,1]}{x-a}\ri)
\eea
We thus have
\bea
\delta_F \le[p_n(x),\wt p_n(x)\ri] =\nonumber \\= \frac 12 \pa_a^k(p_n(a)r_n(a))
[p_n,\wt p_n] +\pa_a^k\frac
{\gamma_n[p_n(a)p_n^\star(a),p_n(a)r_{n\!-\!1}(a)]}{x-a} \mathbf j
\chi_n(x) - \pa_{a}^k \frac {[0,p_n(x)]}{x-a}\label{del1}
\eea
Similarly for the Laurent polynomials
\bea
\delta_F \le[r_{n\!-\!1}(x),\wt r_{n\!-\!1}(x)\ri] =
\e_{n\!-\!1}^t \cdot{\mathbb U^{(\delta_a^k)}}\!^t \cdot  [ \r,\wt\r]\  -\    \e_{n\!-\!1}^t\le(\frac{{\rm d}}{{\rm
    d}z}\ri)^k\bigg|_{z=a} \frac {\r(x)-\r(z)}{x-z}[0,1]\ ,
\eea
where now
\bea
\e_{n\!-\!1}^t \cdot{\mathbb U^{(\delta_a^k)}}\!^t \cdot  [ \r,\wt\r]
=\nonumber \\= -\frac 12 \pa_a^k(p_{n\!-\!1}(a)r_{n\!-\!1}(a))
[r_{n\!-\!1},\wt r_{n\!-\!1}]  +\pa_a^k \e_{n\!-\!1}^t \cdot \r(a)\le(\frac
{\gamma_n[p_n(a),r_{n\!-\!1}^\star (a)]}{x-a} \mathbf j\,{\chi_n^\star}^t(x) -\frac {[0,1]}{x-a}\ri)\label{del2}
\eea
so that finally
\bea
\delta_F \le[r_{n\!-\!1}(x),\wt r_{n\!-\!1}(x)\ri] &=&  -\frac 12 \pa_a^k(p_{n\!-\!1}(a)r_{n\!-\!1}(a))
[r_{n\!-\!1},\wt r_{n\!-\!1}]  +\nonumber \\
&&+\pa_a^k \frac
{\gamma_n[ r_{n\!-\!1}(a)p_n(a), r_{n\!-\!1}(a) r_{n\!-\!1}^\star
    (a)]}{x-a} \mathbf j\,{\chi_n^\star}^t(x)
 - \pa_{a}^k \frac {[0,r_{n\!-\!1}(x)]}{x-a}
\eea
Starifying this last identity and collecting it together with
(\ref{del1}) we finally have
\bea
\delta \chi_n(a) &=& \mathcal U^{(\delta_a^k)}_n(x) \chi_n(x) + \chi_n(x)\mathcal  U^{(\delta_a^k),R}(x)\\
\mathcal  U^{(\delta_a^k)}_n(x) &=& \frac {\pa^k}{\pa a^k}\le\{\frac 1 2\le[\begin{array}{cc}
p_n(a)r_n(a) & 0\\
0& -p_{n\!-\!1}(a)r_{n\!-\!1}(a)
\end{array}\ri] + \frac {\gamma_n}{x-a} \le[\begin{array}{cc}
p_n(a)p_n^\star(a) & p_n(a)r_{n\!-\!1}(a)\\
 r_{n\!-\!1}(a)p_n(a) &  r_{n\!-\!1}(a) r_{n\!-\!1}^\star (a)
\end{array}\ri]\,\mathbf j\ri\}\nonumber \\
&=& \frac {\pa^k}{\pa a^k}  \frac 1 2\le[\begin{array}{cc}
p_nr_n & 0\\
0& -p_{n\!-\!1}r_{n\!-\!1}
\end{array}\ri]_{z=a} + \frac {\pa^k}{\pa a^k}\frac {\gamma_n}{x-a} \le[\begin{array}{cc}
- p_n r_{n\!-\!1} & p_n p_n^\star \\
-  r_{n\!-\!1}  r_{n\!-\!1}^\star & r_{n\!-\!1} p_n
\end{array}\ri]_{z=a}\\
\mathcal  U^{(\delta_a^k),R}(x) &=& \pa_a^k\le[\matrix{0 & \frac 1{a-x}\cr 0&0}\ri]
\eea
This concludes the proof. Q.E.D.\par\vskip 4pt
\section{Moment functionals of integral type  and ODE}
We now assume that the moment functional that we are considering
admits an actual integral representation
\bea
\mathcal L (z^k) := \sum \varkappa_j\int_{\Gamma_j} {\rm
  e}^{-V(z)}z^k{\rm d}z \ .
\eea
As far as the previous discussion on deformations is concerned, the
integral representation of the moment functional is largely
irrelevant, the only issue being the convergence of the deformation
function: therefore the ``potential'' $V(z)$  as well as the sets of integration
$\Gamma_j$ could be  completely
arbitrary. However, in view of  our intentions, we will assume
that $\Gamma_j$ are contours in the complex plane and that $V(z)$ is a
locally defined smooth function on these contours with the only
restriction coming from the fact that negative moments should be
defined as well as the positive ones. \par
 In fact -although many considerations would remain identical in more general situations-  we will
 assume  that $V$ is a locally analytic function in the complex
$z$-plane excepted at some punctures, identically to the case of
{\bf semiclassical moment functionals} studied in \cite{B,semiiso} 
with the only extra restriction that all negative moments should be
defined and finite.
\paragraph{Semiclassical Moment Functionals.}
For the reader's convenience we briefly recall how these semiclassical
 moment functionals are constructed \cite{B,semiiso,MR1,MR2}.
 In this case the potential is such that the derivative is an arbitrary rational function 
 \be
 V'(z) = \hbox{Rational function}
 \ee
 and thus $V(z)$ is a rational function plus logarithmic singularities at those poles of $V'$ where the residue does not vanish. For simplicity we assume that $V'$ has either a pole or a nonzero limit at $z=\infty$.
 Once we have chosen the potential $V$ we also choose an arbitrary  collection of   contours (avoiding $z=0$) $\{\Gamma_j\}$ with the property that  $\Re(V(x))$ is uniformly bounded from below on all the chosen contours and tends to $\infty$ polynomially (in the length parameter) on the contours that extend to $z=\infty$. In more detailed terms:
 \begin{enumerate}
 \item[{\bf (a)}] Consider a pole $z=c$ of $V'$ of order $k\geq 2$: we attach to it $k-1$ ``petals'' approaching $z=c$ along asymptotic directions in the sectors where $\Re(V(x))\to +\infty$. We also attach a ``stem'' extending to $\infty$ and asymptotic to a direction such that $\Re(V(x))\to \infty$.  
\item[{\bf (b)}] For a simple pole $z=c$ of $V'$, if the residue is a positive integer (i.e. ${\rm e}^{-V}$ has a pole at $z=c$) we choose a small loop around the point, if the residue is a negative integer we take a contour from $z=c$ to $\infty$, if the residue is non integer we take a loop coming from $\infty$ and returning to $\infty$  (with the same restriction as above for the asymptotic direction).
\item[{\bf (c)}] We choose also arbitrary segments joining a certain number of points $z=a$  to $\infty$ (along admissible directions). These latter contours are called ``hard-edge'' contours because the pseudo measure ${\rm d}\mu = {\rm e}^{-V(z)} {\rm d}z$  has a limit at $z=a$ and integration by parts yields a boundary term.
\end{enumerate}
\begin{center}
\epsfxsize 8cm
\epsfysize 8cm
\epsffile{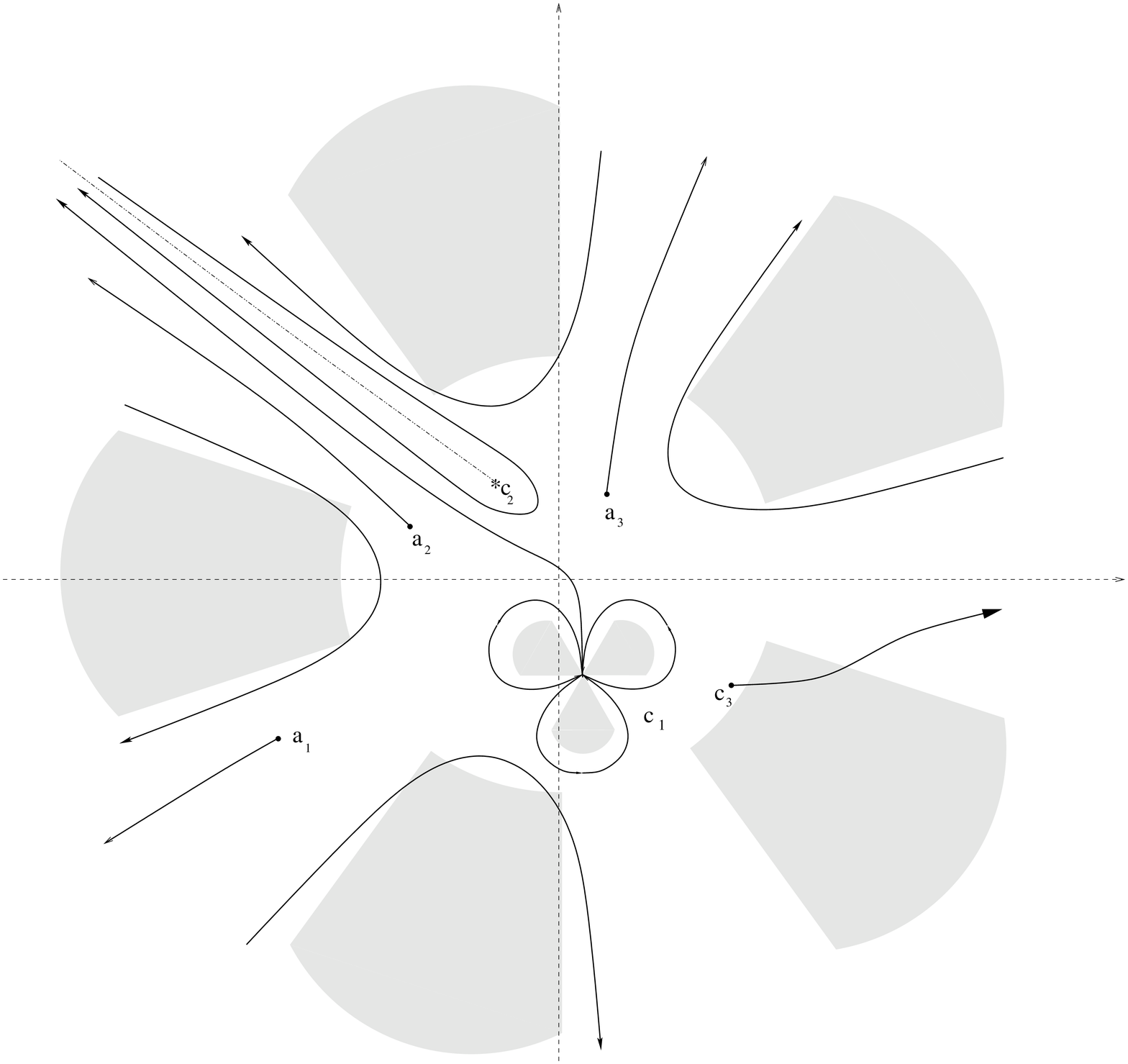}
\parbox{12cm}{Figure: The contours for a typical semiclassical moment functional. Here $V'(x)$ has a pole of order $4$ at $\infty$, of order $4$ at $c_3$ and  simple poles at $c_2,c_3$ with noninteger and negative-integer residue respectively. The contours originating from the $a_i $'s are ``hard-edge'' contours. The shaded sectors represent the asymptotic ``forbidden'' directions for approaching a singularity. One of these sectors at $\infty$ in Figure does not have a contour surrounding it because such a  contour would be ``homologically'' equivalent to minus the sum of all others.}
\end{center}

\subsection{Differential equations}
We first analyze in this situation the infinite-dimensional
differential equation that the BOPs satisfy.
The natural differential operation in this setting is not $\pa_x$ but
rather $x\pa_x$.  Using the recurrence relations involving
multiplication by $x$ and the orthogonality relations
\bea
&& x \p(x) = Q \p(x)\ ;\ x \r^t(x) = \r^t(x) Q\\
&& x\p'(x) = D\p(x)\ ;\ x{\r}'(x) = \tilde  D\r(x)\\
 &&\int_\varkappa \p \r^t {\rm
e}^{-V}{\rm d}z = \1\ ,
\eea
we can obtain the following identity by integrating $\pa_z(z\p\r^t{\rm e}^{-V(z)})$:
\bea
&& D + {\tilde D}^t - \le(z\p(z)\r^t(z){\rm e}^{-V(z)}\ri)\Big|_{\pa\varkappa}  =Q V'(Q)- \1
 \eea
The notation that we now adopt is that $\int_\varkappa$ stands for the linear
combination with coefficients $\varkappa_j$ of integrals on the
oriented contours $\Gamma_j$ and the evaluations $\big|_{\pa
  \varkappa}$ stand for the evaluations at all endpoints of the given
contours, multiplied by the corresponding coefficient $\varkappa$ and
the appropriate sign according to the orientation.
 The matrices
$D$ and $\tilde D$ are lower triangular and on the main diagonal they
 can be explicitly computed
 \bea
 &&xp_n' = np_n + \hbox{ previous };\\
&& xr_n' =
\le((-n)(1-\dot\ell_{n+1})+\ell_{n+1}\ri) r_n + \hbox{ previous }\\
 &&
x\frac{{\rm d}}{{\rm d}x} r_{n}^\star = n\dot\ell_{n+1}r_n^\star +
\hbox{ previous }\ .
  \eea
 This implies the following Virasoro scaling constraint
\be
(QV'(Q))_{nn}+ (zp_n\,r_n {\rm
  e}^{-V})\bigg|_{\pa\varkappa} =1
+\ell_{n+1} +n\dot\ell_{n+1} \label{diagscal}\ .
\ee
Note that we also have
\be
\sum_{k=0}^{n-1}\le( (QV'(Q))_{kk }+ (zp_k\,r_k {\rm
  e}^{-V})\bigg|_{\pa\varkappa}\ri) = \sum_{k=0}^{n-1} \le(1
+\ell_{l+1} +l\dot\ell_{k+1}\ri) = n(\ell_n+1) \label{tracevir}\ .
\ee
The parts of $D, \wt D$ below the main diagonal are now expressed in
terms of $Q$ and the boundary terms only
\bea
D_< = (QV'(Q))_< + \le(z(\p(z)\r^t(z))_<{\rm
  e}^{-V(z)}\ri)\Big|_{\pa\varkappa} \ ;\\
\wt D_< = (Q^tV'(Q^t))_< + \le(z(\r(z)\p^t(z))_<{\rm
  e}^{-V(z)}\ri)\Big|_{\pa\varkappa}
\eea
Note that -below the main diagonal- the matrices $D$ and $\wt D$ are of
the same form as the deformations we were considering previously; more
precisely they correspond to a variation by $ F(z) = zV'(z)$ and a linear
combination of variations supported at the endpoints of the contours
$\Gamma_j$.
The folded version of this ODE can be obtained from the formul\ae\
(\ref{def1}, \ref{def2}, \ref{del1}, \ref{del2}) with the only
modification that comes from the diagonal part of $D$. Using
(\ref{diagscal}) for the diagonal part the reader can check that the result is
\bea
\hspace{-2cm}\DD_n &=& \le[\matrix {n & 0 \cr
0 & xV'(x)-1-\ell_n} \ri] +\gamma_n \le[\matrix{-W_{n,n-1} & W_{n,n^\star}\cr -W_{(n-1)^\star,n-1} &
W_{n,n-1}}\ri]  +\le(\frac {z{\rm e}^{-V(z)} \gamma_n}{x-z} \le[\begin{array}{cc}
- p_n r_{n\!-\!1} & p_n p_n^\star \\
-  r_{n\!-\!1}  r_{n\!-\!1}^\star & r_{n\!-\!1} p_n
\end{array}\ri]\ri)\Bigg|_{\pa\varkappa} \nonumber\\[20pt]
 && W := \nabla_QxV'(x) = \frac {QV'(Q)-xV'(x)}{Q-x}\label{DDform}
\eea
We remark that the last ``boundary'' term consists of  simple poles with
nilpotent residues.\par
For the full matrix $\chi_n$ the differential equation is
\bea
x\pa_x \chi_n(x) &=& \DD_n(x)\chi_n(x) + \chi_n(x)\DD^{R}(x)\\[20pt]
\DD^R(x) &=& \le[\begin{array}{cc}
0 & \ds\int_\varkappa \frac {xV'(x)-zV'(z)}{x-z}{\rm e}^{-V(z)} {\rm
  d}z + \frac {z{\rm e}^{-V(z)}}{z-x}\bigg|_{\pa\varkappa} \\[10pt]
0 & -xV'(x)
\end{array}\ri]
\eea
Together with the differential equation and the deformation equations
we recall that we also have  difference equations
\bea
\chi_n &= & R_n(x) \chi_{n-1}\ ,\qquad n\geq 1
\\
 R_n(x)&=&   \le\{
\begin{array}{cl}
\ds \le[\begin{array}{cc}
\frac {x-\beta_n}{\gamma_n} & \kappa_n
\\
(-1)^{n+1} & 0
\end{array}\ri] &\hbox { if } \dot\ell_n = 1
\\[20pt]
\ds \le[\begin{array}{cc}
\frac {x-\beta_n}{\gamma_n} & \kappa_n
\\ (-1)^{n+1} & \omega_n
\end{array}\ri] &\hbox { if } \dot\ell_n = 0
\end{array}\ri.
\eea
The ladder matrices $R_n$ are simply obtained from the transfer
matrices (\ref{transfer}) by using the normalization of the
polynomials as in (\ref{pirho}).
We have thus  proved
\bt
\label{ciccio}
The matrix $\chi_n$ satisfies the following system of difference-deformation-differential ({\em DDD}
for short) equations
\bea
\chi_n &= & R_n(x) \chi_{n-1}\\
x\frac{\rm d}{{\rm d}x} \chi_n &=& \DD_n \chi_n + \chi_n \DD^R\\
\delta_f \chi_n &=& \mathcal U_n^{(f)} \chi_n  + \chi_n  \mathcal U^{(f),R}
\eea
where $f$ denotes either any function or formal power series provided that
$\mathcal L(f(z)z^k)$ is well defined for $k\in \Z$ or any derivative
of the Dirac delta function supported at any point $a\neq 0$.
\et
We observe that the right action of the differential-deformation
equation is independent of $n$. This suggests that we can perform a
``right gauge'' change to dispose of this part.
Indeed we define the new object $\GG_n$ which will be the focus in the
rest of the paper
\be
\GG_n:= \chi_n \le[\begin{array}{cc}
1 & \ds -{\rm e}^{V(x)} \int_\varkappa \frac{{\rm e}^{-V(z)}}{x-z}{\rm
  d}z\\[10pt]
0 & {\rm e}^{V(x)}
\end{array}\ri]
\ee
It is easy to verify that this change of gauge eliminates the
right-actions for the differential equation and for any deformation of
$V(x)$ and/or the endpoints of integration. The first column of
$\GG_n$ is the same as the first column of $\chi_n$ and hence contains
the LOPs. The second column contains now  the following {\bf auxiliary
  functions}
\bea
\psi_n &=& {\rm e}^{V(x)}\int_\varkappa \frac {p_n(z) {\rm e}^{-V(z)}}
    {x-z}{\rm d}z\\
\phi_{n-1}^\star &=&x^{n-1-\ell_n}\phi_{n-1} = x^{n-1-\ell_n} {\rm
  e}^{V(x)}\int_\varkappa \frac {r_{n-1}(z) {\rm e}^{-V(z)}}
    {x-z}{\rm d}z\ .
\eea
We note that the auxiliary functions are piecewise analytic functions
off the contours $\Gamma_j$: it is a matter of routine inspection to
read-off the relevant Riemann-Hilbert data. We defer this inspection
to a later section. \par
In terms of the matrices $\GG_n$ we have a DDD system of more standard
form, without right multipliers.
\bt
\label{cicciobello}
The following system of Difference-Differential-Deformation equations
is Frobenius compatible
\bea
\GG_n &= & R_n(x) \GG_{n-1}\\
x\frac{\rm d}{{\rm d}x} \GG_n &=& \DD_n \GG_n\\
\delta_f \GG_n &=& \mathcal U_n^{(f)} \GG_n
\eea
where $f$ is as in Thm. (\ref{ciccio}).
\et
A few remark are in order here: by choosing $f$ in
Thm. (\ref{cicciobello}) to be an ordinary function one can
vary the potential $V$ by $V\to V+\epsilon f$ and hence all flows of
the generalized Toda hierarchy are here included. However we can also
choose $f$ as a distribution $\delta_a^{(k)}$ or linear combinations
thereof. Clearly if we choose the point $a$ arbitrarily outside of the
singularities of $V(x)$ we still have
a compatibility of the resulting system but we will change the
structure of the singularities of $\DD_n$, which falls outside of the
standard theory of isomonodromic deformations. For example, adding a
$\delta_a$ corresponds to adding a term $\ln(x-a)$ in the potential
{\bf and} adjoining a small circle around $a$ to  the set of contours
$\Gamma_j$'s.\\
Vice-versa the cases in which $f$ is a distribution which does not
alter the singularity structure of $\DD_n$ are:
\begin{enumerate}
\item Movement of the endpoints which {\bf contribute} to the
  boundary term\footnote{They corresponds to those endpoints of the
  contours $\Gamma_j$ for which $\ds \lim_{\Gamma_j \ni z\to \pa \Gamma_j} {\rm
  e}^{-V(z)}\neq 0$.}: then we have
\be
f = \pm\varkappa {\rm e}^{-V(a)} \delta_{a}
\ee
where the coefficient $\varkappa$ is the coefficient of the contour
  $\Gamma_j$ which has $a$ as endpoint and the sign depends on the
  orientation of $\Gamma_j$.
\item Movements of {\bf poles} of order $k$ (if any) of the pseudo-measure ${\rm e}^{-V}{\rm
  d}z$: then we have
\be
f(a) = \pm \varkappa\,k\, \delta_{a}^{(k+1)}(z){\rm e}^{-V_r(z)}
\ee
where the coefficient $\varkappa$ is the coefficient of the loop
  encircling $a$, the sign is chosen according to the orientation
  of the contour and $V_r(z)$ is the part of $V$ which is regular at $z=a$.
\end{enumerate}
\section{Spectral curve and Isomonodromic tau function}
The objective of this section is that of expressing the spectral curve of the connection $\pa_x - \frac 1  x \DD_n(x)$ in terms of the logarithmic derivatives of the T\"oplitz determinants; this will be the essential bridge to connect with the isomonodromic tau function in the coming sections.
We prove the following theorem
\bt
The following formula holds
\bea
\det\le(y\1-\frac 1 x \mathcal D_n(x)\ri)& =& y^2 - \le(V'(x) +\frac{L_n}x\ri)y +\nonumber \\
&& +
\frac 1 x \tr_n\le(\frac {QV'(Q)-xV'(x)}{Q-x}  \ri) +\frac 1 x
\le(\frac {z{\rm e}^{-V(z)}\p^t\Pi_{n-1}\r }{x-z}
\ri)\Bigg|_{\pa\varkappa}  \\
&& L_n:=n-1-\ell_n\ ,
\eea
where $\Pi_{n-1} = {\rm diag}(1,1,\dots,1,0,\dots)$ ($n$ nonzero entries).
\et
Before proceeding to the proof we remark that this formula would be valid for an arbitrary {\em smooth}  potential; quite clearly, however, in this case the spectral curve would not be an algebraic curve.\\
{\bf Proof}.\\
We need to compute the two spectral invariants of the connection; the main tool is to use the compatibility between the ladder relations and the connections $\DD_n(x)$. Indeed
from the compatibility between the difference and differential
equation and from the explicit expression for $\DD_n(x)$ (\ref{DDform})
 we can express recurrence relation for the spectral invariants of
 $\DD_n(x)$. The trace is computed by sight
\bea
 \tr(\mathcal D_n(x))
&=& xV'(x)+n-1-\ell_{n}
\eea
From the compatibility of difference-differential equations we have
the gauge property
\bea
 \mathcal D_{n-1} &=& { R_{n}}^{-1} \mathcal
D_n R_n -x\, {R_n}^{-1} R_n'
 \eea
The gauge term is explicitly computed to be
\bea
{ R_{n}}^{-1} \mathcal D_n R_n  &=&\mathcal D_{n-1} + x\,{R_n}^{-1}
R_n'\\
x\,{R_n}^{-1} R_n' &=& \le\{\begin{array}{cl}
\ds (-1)^n\frac x {\gamma_{n-1}}\le[\matrix{0 &0\cr 1& 0}\ri] & \hbox
    { if }\dot\ell_n=1\\[20pt]
\ds \le[\matrix{1 &0\cr \frac{(-1)^n
      \Delta_{n-1}^{\ell_n-1}}{\Delta_n^{\ell_n}
      \sqrt{\Delta_{n-2}^{\ell_{n-2}}}} &  0}\ri] & \hbox
    { if }\dot\ell_n=0
\end{array}\ri.
\eea
These formul\ae\ imply a recurrence relation for the quadratic
invariant.
\bea
\tr (\mathcal D_{n}^2)) &=&  \tr(\mathcal D_{n-1}^2) + 2\tr\le( \mathcal
D_{n-1} x{R_n}^{-1} R'_n \ri) +\tr((x{R_n}^{-1} R_n')^2).
\eea

For the {\em line case} i.e. $\dot\ell_n=1$ and using the form of the
recursion matrices $R_n$ together with the fact that in this case
$r_{n-1}^\star = (-1)^{n-1}p_{n-1}$ we find
\bea
\tr(\mathcal D_{n}^2)&=&  \tr(\mathcal D_{n-1}^2) - 2x
\le(\frac{QV'(Q)-xV'(x)}{Q-x}\ri)_{n-1,n-1}
-2x \le(\frac{ z{\rm e}^{-V(z)p_{n\!-\!1}r_{n\!-\!1} }} {x-z}\ri)\bigg|_{\pa\varkappa}
 \label{recsumline}
\eea
For the circle case $\dot\ell_n=0$ instead we have
\bea
\tr(\mathcal D_{n}^2)&=&  \tr(\mathcal D_{n-1}^2) + 2(n-1) +
2\gamma_{n-1}\le(- W_{n-1,n-2} +  \frac{(-1)^n
      \Delta_{n-1}^{\ell_n-1}}{\Delta_n^{\ell_n}
      \sqrt{\Delta_{n-2}^{\ell_{n-2}}}} W_{n-1,(n-1)^\star}\ri)
+\nonumber \\
&&+ 2\gamma_{n-1} \le(\frac {z{\rm e}^{-V(z)} p_{n-1}} {x-z}
\le(-r_{n-2} +   \frac{(-1)^n
      \Delta_{n-1}^{\ell_n-1}}{\Delta_n^{\ell_n}
      \sqrt{\Delta_{n-2}^{\ell_{n-2}}}} p_{n-1}^\star\ri)\ri)\bigg|_{\pa\varkappa}+ 1 \label{recsumcirc}
\eea
Using the identity (\ref{uno}) together with the definitions of the
biorthogonal polynomials and the various normalization factors
(\ref{pirho}) one can see that
\be
-r_{n-2} +   \frac{(-1)^n
      \Delta_{n-1}^{\ell_n-1}}{\Delta_n^{\ell_n}
      \sqrt{\Delta_{n-2}^{\ell_{n-2}}}} p_{n-1}^\star = -\frac
{z}{\gamma_{n-1}} r_{n-1}\ ,
\ee
and hence
\bea
-W_{n-1,n-2} +   \frac{(-1)^n
      \Delta_{n-1}^{\ell_n-1}}{\Delta_n^{\ell_n}
      \sqrt{\Delta_{n-2}^{\ell_{n-2}}}} W_{n-1,(n-1)^\star} =\frac 1 {\gamma_{n-1}}
\mathcal L_z\Bigg(z\,W p_{n-1}
  r_{n-1}\Bigg)
\eea
Therefore the recursion for the circle case is
\bea
\tr(\mathcal D_{n}^2)-\tr(\mathcal D_{n-1}^2)&=&   2(n-1) + 1
-2\le(\frac{ Q\le(QV'(Q)-xV'(x)\ri)} {Q-x}\ri)_{n-1,n-1} \hspace{-20pt}-2 \le(\frac
{z^2{\rm e}^{-V(z)}p_{n-1}r_{n-1}}{x-z}\ri)\Bigg|_{\pa\varkappa}=
\nonumber \\
&=&  2(n-1) + 1 -2 \le(QV'(Q)_{n\!-\!1,n\!-\!1} + (z{\rm e}^{-V(z)}
p_{n\!-\!1}r_{n\!-\!1})|_{\pa\varkappa}\ri) +2xV'(x)+\nonumber \\
&& -2x\le(\frac{QV'(Q)-xV'(x)} {Q-x}\ri)_{n-1,n-1} - 2x \le(\frac
{z{\rm e}^{-V(z)}p_{n-1}r_{n-1}}{x-z}\ri)\Bigg|_{\pa\varkappa}
=\nonumber \\
&=&  2(n-1)- 1 - 2\ell_{n}  +2xV'(x)
-2x\le(\frac{QV'(Q)-xV'(x)} {Q-x}\ri)_{n-1,n-1}\hspace{-20pt} +
\nonumber \\
&& - 2x \le(\frac
{z{\rm e}^{-V(z)}p_{n-1}r_{n-1}}{x-z}\ri)\Bigg|_{\pa\varkappa}
\eea
Summarizing, in the two cases we have  found
\bea
\tr(\mathcal D_{n}^2)-\tr(\mathcal D_{n-1}^2)&=&
2\le(xV'(x)- \ell_{n} +(n-1) -\frac 1 2 \ri)(1-\dot \ell_n)+\nonumber \\
&&
-2x\le(\frac{QV'(Q)-xV'(x)} {Q-x}\ri)_{n-1,n-1}\hspace{-18pt}- 2x \le(\frac
{z{\rm e}^{-V(z)}p_{n-1}r_{n-1}}{x-z}\ri)\Bigg|_{\pa\varkappa}
\eea
To complete the computation we need to find $\tr({\mathcal D_1}^2)$ or
--equivalently-- $\det(\mathcal D_1)$.
We have
\bea
\det\le(\frac 1  x D_1\ri) &=& \det\le[\matrix{p_1' & \psi_1' \cr {r_0^\star}' &
    {\phi_0^\star}'}\ri] \le[\matrix{p_1 & \psi_1 \cr {r_0^\star} &
    {\phi_0^\star}}\ri]^{-1} =  \det\le[\matrix{p_1' & \psi_1' \cr {r_0^\star}' &
    {\phi_0^\star}'}\ri]{\rm e}^{-V(x)} =\nonumber \\
&=&\sqrt{h_1}{\rm e}^{-V(x)}
\det\le[\matrix{\frac 1 {\sqrt{h_1}}  & \psi_1' \cr 0  &
    {\phi_0^\star}'}\ri] = \nonumber \\
&=& V'(x)\mathcal L_z\le(\frac 1 {x-z}\ri) -\mathcal
L_z\le(\frac{V'(z)} {x-z}\ri)  + \le(\frac {{\rm e}^{-V(z)}}{x-z}
\ri)\Bigg|_{\pa\varkappa} =\nonumber \\
&=& \le(\frac{V'(Q)-V'(x)}{Q-x}\ri)_{00} + \le(\frac {{\rm e}^{-V(z)}p_0r_0}{x-z}
\ri)\Bigg|_{\pa\varkappa}
\eea
This implies
\bea
\det \mathcal D_1(x) &=& x^2 \le(\frac{V'(Q)-V'(x)}{Q-x}\ri)_{00} +x^2 \le(\frac {{\rm e}^{-V(z)}p_0r_0}{x-z}
\ri)\Bigg|_{\pa\varkappa}    =\nonumber \\
&=& \hspace{-4pt} x\overbrace{\le(V'(Q)_{00} + (p_0r_0{\rm e}^{-V(z)})|_{\pa\varkappa} \ri)}^{=0} + x
\le(\frac{QV'(Q)-xV'(x)}{Q-x}\ri)_{00} \hspace{-5pt} +x \le(\frac {z{\rm e}^{-V(z)}p_0r_0}{x-z}
\ri)\Bigg|_{\pa\varkappa}   \hspace{-5pt}=\nonumber \\
&=&  x
\le(\frac{QV'(Q)-xV'(x)}{Q-x}\ri)_{00}   +x \le(\frac {z{\rm e}^{-V(z)}p_0r_0}{x-z}
\ri)\Bigg|_{\pa\varkappa}
\eea
Hence ($\ell_1=0$)
\bea
&& \tr({\mathcal D_1}^2) = \le(xV'(x)\ri)^2
-2x\le(\frac{QV'(Q)-xV'(x)}{Q-x}\ri)_{00}  -2x \le(\frac {z{\rm e}^{-V(z)}p_0r_0}{x-z}
\ri)\Bigg|_{\pa\varkappa} \\
&& \tr({\mathcal D_n}^2) = (xV'(x))^2 + 2 xV'(x)\le(n-1-\ell_n\ri) -
(n-1-\ell_n) + 2\sum_{k=1}^{n} (k-1-\ell_{k})(1-\dot \ell_k)
+\nonumber \\
&& \hspace{2cm} -2x\tr_n\le(\frac {QV'(Q)-xV'(x)}{Q-x}\ri)  -2x \le(\frac {z{\rm e}^{-V(z)}\p^t\Pi_{N-1}\r}{x-z}
\ri)\Bigg|_{\pa\varkappa}  \label{trD2}\ .
\eea
Using this expression for the quadratic invariant we can obtain the
following formula for the characteristic polynomial
\bea
\det(\tilde y\1-\mathcal D_n(x))& =& \tilde y^2 - \big(xV'(x) +n-1-\ell_n\big)\tilde y +
K_n +\nonumber \\
&& +
x\tr_n\le(\frac {QV'(Q)-xV'(x)}{Q-x}  \ri) +x \le(\frac {z{\rm
    e}^{-V(z)}\p^t\Pi_{N-1}\r }{x-z}
\ri)\Bigg|_{\pa\varkappa}\nonumber  \\
&&K_n:=\frac{(n-1-\ell_n)(n-\ell_n)}2 + \sum_{k=2}^{n}
(\ell_k+1-k)(1-\dot\ell_k) \label{speccurv}
\eea
The last crucial observation is that $K_n \equiv 0$ for all $n$: this
is non-obvious at first sight and it is true only because $\ell_n$ is
a weakly increasing sequence of {\em integers}. Indeed one can check
that
\be
K_{n+1}- K_n = \frac 12 \dot \ell_{n+1}(1-\dot \ell_{n+1})\ ,
\ee
so that $K_{n+1}=K_n = K_1=0$. To conclude the proof we note that the spectral curve of (\ref{speccurv}) is simply related to that of the connection by $\tilde y = x y$. This ends our proof.

\section{Isomonodromic deformations}
By Thm. \ref{cicciobello} we have compatible systems of Difference, Differential, Deformation
equations
\bea
\GG_n &=& R_n \GG_{n-1}\\
\pa_x \GG_n &=& \frac 1 x \mathcal D_n \GG_n\\
\delta_f \GG_n &=& \mathcal U^{(f)}_n(x) \GG_n
\eea
The compatibility of this system entails isomonodromic deformations
for the connection $\pa_x - \frac 1 x \mathcal D_n$.
Note that this connection has the same singularity structure of
$V'(x)$. In order to have isomonodromic deformations in the sense of
Miwa-Jimbo-Ueno we need to impose that $V'(x)$ be a rational
function. Then the deformations of $V(x)$ which give rise to the
setting in MJU are those which do not alter the singularity structure
of $V(x)$; this is why the most general setting compatible with this requirement is that of semiclassical moment functionals.
\subsection{Spectral residue-formul\ae}
Mimicking the approach of \cite{semiiso} we can express the
logarithmic derivatives of the shifted T\"oplitz determinants
$\Delta_{n}^{\ell_n}$ in terms of residue formul\ae\ involving the
differential $y{\rm d}x$ on the spectral curve defined in
Thm. \ref{cicciobello}
\bea
Y_\pm(x)&\!\!\!:=&\!\! \frac 1 2 \le(V'(x) +\frac{L_n}x\ri) \pm \frac 1 2\sqrt{
  \le(V'(x) +\frac{L_n}x\ri) ^2 - 4 \mathcal P(x)}\\
\mathcal P(x) &\!\!\!\!\!=&
\frac 1 x \tr_n\le(\frac {QV'(Q)-xV'(x)}{Q-x}  \ri) +\frac 1 x
\le(\frac {z{\rm e}^{-V(z)}\p^t\Pi_{N-1}\r }{x-z}
\ri)\Bigg|_{\pa\varkappa}
\eea
Indeed we have
\bt  Let $V'(x)$ be rational.\\
{\bf (i)} Suppose that $x=c$ is a pole of order
$d+1$
\bea
V(x) = \sum_{J=1}^{d} \frac{t_J^{(c)}}{J\,(x-c)^J} - t_0^{(c)} \ln(x-c) + \mathcal
O(1)\\
V'(x) = -\sum_{J=0}^{d} \frac{t_J^{(c)}}{(x-c)^{J+1}}  + \mathcal
O(1)
\eea
Then we have
\bea
t_J^{(c)} &=&- \res{x=c} Y_+(x) (x-c)^{J}{\rm d}x\ ,\qquad J=0,\dots d \label{tc}\\
\frac {\pa \ln \Delta_n^{\ell_n}}{\pa t_J^{(c)}}& =& \frac 1 J\res{x=c} Y_-(x) (x-c)^{-J}{\rm
  d}x\ ,\qquad J=1,\dots, d\\
\frac {\pa \ln \Delta_n^{\ell_n}}{\pa c}& =& \res{x=c} Y_-(x)
\le(\sum_{J=0}^{d} \frac{t_J^{(c)}}{(x-c)^{J+1}}\ri){\rm d} x
\eea
{\bf (ii)} Suppose that $x=\infty$ is a pole of $V'$ with degree
$d$, namely
\bea
V(x) = \sum_{J=1}^{d+1} \frac {t_J^{(\infty)}}J x^J + \mathcal O(\ln x)\\
V'(x) = \sum_{J=1}^{d+1} {t_J^{(\infty)}} x^{J-1} + \mathcal O(1/ x)
\eea
Then we have
\bea
t_J^{(\infty)} &=& -\res{x=\infty} Y_+(x) x^{-J}{\rm d}x\ ,\qquad J=1,\dots
d+1\label{tinf} \\
\frac {\pa \ln \Delta_n^{\ell_n}}{\pa t_J^{(\infty)}}& =& \frac 1 J\res{x=\infty} x^J
Y_-(x){\rm d}x\ ,\qquad J=1,\dots,d+1
\eea
{\bf (iii)} Let $x=a$ be a {\bf hard-edge}\footnote{This means that
  this is one of the points contributing to the boundary terms.}, namely a point of the
boundary of one of the contours $\{\Gamma_j\}$ such that
$|V(a)|<\infty$. Then
\bea
\frac {\pa \ln \Delta_n^{\ell_n}}{\pa a}& =&  \frac 12 \res{x=a} \frac 1 {x^2}\tr
(\DD_n)^2{\rm d}x
\eea
{\bf (iv)} Finally we have
\bea
&&\res{x=0}Y_+(x){\rm d}x = L_n = n-1-\ell_n - \sum_{c\hbox{\small =
    finite pole of } V'} t_0^{(c)} \nonumber\\
&&\res{x=\infty} Y_+(x){\rm d}x = \ell_n+1 + t_0^{(\infty)} \label{schles}
\eea
\et
{\bf Proof.}
We start by noticing that
\be
Y_\pm = \le\{{1\atop 0} \ri\} \le( V'(x) + \frac {L_n}x \ri) \mp
\frac {\mathcal P(x)}{V'(x)+\frac {L_n}x } + \le\{
\begin{array}{cl}
\mathcal O((x-c)^{d+1}) & \hbox { for case {\bf (i)}} \\[15pt]
\ds \mp \frac {n^2}{t_{d+1}^{(\infty)} x^{d+1}} +\mathcal O(x^{-d-2}) &  \hbox{
  for case {\bf (ii)}}
\end{array}
\ri.\ \label{Yas}
\ee
At this point formul\ae\ (\ref{tc},\ref{tinf} ,\ref{schles}) follow immediately by
noticing that $\mathcal P/(V'(x)+L_n/x) = \mathcal O(1)$ in all
cases and by straightforward computation of residues\footnote{ Note
  that at infinity
\be
\frac {\mathcal P(x)}{V'(x)+\frac {L_n}x } = \frac n x + \mathcal
O(x^{-2})\ .
\ee
}.
As for the remaining formul\ae\ we have, for case {\bf (i)}
\bea
&&\hspace{-2.5cm} \res{x=c} (x-c)^{-J} Y_-(x){\rm d}x =  \res{x=c}
(x-c)^{-J}  \frac {\mathcal P(x)}{V'(x)+\frac{ L_n}x } =
\res{x=c} \frac { (x-c)^{-J}} {xV'(x)+L_n  }  \tr_n \le(\frac
    {xV'(x)-QV'(Q)}{x-Q}\ri) = \nonumber \\
&&\hspace{-2cm}  = \sum_{k=0}^{n-1}\mathcal L_z\le[ \res{x=c} \frac { (x-c)^{-J}} {xV'(x)+{L_n} } \frac
    {xV'(x)-zV'(z)}{x-z} p_n(z)r_n(z) \ri] =\nonumber \\[10pt]
&&=   -\tr_n (Q-c)^{-J} =J\pa_{t_J^{(c)}} \ln \Delta_{n}^{\ell_{n}}\ , \
\ J=1,\dots, d\ ,
\eea
and similar computation for the $c$-derivative. Here we have used the
formul\ae\ (\ref{defDelta}) expressing the variation of $\ln
\Delta_n^{\ell_n}$ under an infinitesimal deformation of the type
ensuing from an infinitesimal change of the parameters $t_J^{(c)}$.\\
For case {\bf (ii)} the computation is completely parallel except for
the last $J=d+1$ residue. Indeed
\bea
&&\hspace{-2.5cm} \res{x=\infty} x^{J} Y_-(x){\rm d}x =  \res{x=\infty} x^{J}  \frac {\mathcal P(x)}{V'(x)+\frac {L_n}x } =
\res{x=\infty } \frac { x^{J}} {xV'(x)+L_n }  \tr_n \le(\frac
    {xV'(x)-QV'(Q)}{x-Q}\ri) = \nonumber \\
&&\hspace{-2cm}  = \sum_{k=0}^{n-1}\mathcal L_z\le[ \res{x=\infty} \frac { x^{J}} {xV'(x)+L_n} \frac
    {xV'(x)-zV'(z)}{x-z} p_n(z)r_n(z) \ri] =\nonumber \\[10pt]
&&=  -\tr_n Q^{J} =J\pa_{t_J^{(\infty)}} \ln \Delta_{n}^{\ell_{n}}\ , \
\ J=1,\dots, d\ .\label{lessJ}
\eea
For $J=d+1$ one has to use a similar manipulation but has to use the
refined asymptotics (\ref{Yas}): indeed we have
\bea
 \res{x=\infty} xV'(x) Y_-(x){\rm d}x =- \tr_n QV'(Q) + \le(n^2-nL_n
 -\tr_n QV'(Q) - z{\rm e}^{-V(z)} \p^t\Pi_{n-1} \r
 \bigg|_{\pa\varkappa}\ri) =\nonumber \\
=  -\tr_n QV'(Q)
\eea
where we have used (\ref{tracevir}) together with the definition of
$L_n = n-1-\ell_n$. This proves, together with the residues (\ref{lessJ})
\be
\res{x=\infty}x^{d+1} Y_-(x){\rm d}x = -\tr_n Q^{d+1} =
(d+1)\pa_{t_{d+1}^{(\infty)}} \ln \Delta_n^{\ell_n}\ .
\ee
Finally, for the case {\bf (iii)}  the computation is immediate using
the formula for $\tr \DD_n^2$ (\ref{trD2}). Q. E. D. \par \vskip 4pt
\section{Riemann--Hilbert problem, Tau function}
Direct inspection of the asymptotic behavior of the biorthogonal
polynomials and second kind functions allows us to ascertain the
Riemann--Hilbert
data for this problem.
We start by noticing the following formal asymptotic behavior of the
auxiliary functions entering in $\GG_n$
\bea
\psi_n = {\rm e}^{V(x)}\int_{\varkappa} \frac {{\rm
    e}^{-V(z)}p_n(z)}{x-z} &=& \le\{
\begin{array}{ll}
\ds (-)^n x^{-\ell_n-2} {\rm e}^{V(x)} \sqrt {\frac
    {\Delta_{n+1}^{\ell_{n+1}}}{\Delta_n^{\ell_n}}} (1+ \mathcal
    O(x^{-1})) & \hbox { for } x\to \infty\\[10pt]
\ds -x^{n-1-\ell_n}{\rm e}^{V(x)} \frac
    {\Delta_{n+1}^{\ell_n}}{\sqrt{\Delta_n^{\ell_n}\Delta_{n+1}^{\ell_{n+1}}}}
    (1+\mathcal O(x)) & \hbox { for } x \to 0\\[10pt]
\ds {\rm e}^{V(x)} \sqrt{h_0} (Q-c)^{-1}_{n0} & \hbox {near poles of }V'(x)
\end{array}
\ri.\\[20pt]
\phi_{n\!-\!1}^\star  = x^{n-1-\ell_n} {\rm e}^{V(x)} \int_{\varkappa} \frac {{\rm
    e}^{-V(z)}r_{n\!-\!1}(z)}{x-z} &=& \le\{
\begin{array}{ll}
\ds  x^{-\ell_n-1} {\rm e}^{V(x)} \sqrt {\frac
    {\Delta_{n}^{\ell_{n}}}{\Delta_{n-1}^{\ell_{n-1}}}} (1+ \mathcal
    O(x^{-1})) & \hbox { for } x\to \infty\\[10pt]
\ds x^{n-1-\ell_n}{\rm e}^{V(x)} \frac
    {(-)^n \Delta_{n}^{\ell_n-1 }}{\sqrt{\Delta_n^{\ell_n}\Delta_{n-1}^{\ell_{n-1}}}}
    (1+\mathcal O(x)) & \hbox { for } x \to 0\\[10pt]
\ds {\rm e}^{V(x)} \sqrt{h_0} c^{n-\ell_n+1} (Q-c)^{-1}_{0,n\!-\!1} & \hbox {near poles of }V'(x)
\end{array}
\ri.
\eea
where we have used the definition of the LOPs (\ref{pirho}) and the
facts that
\bea
p_n\propto\wp_n^{\ell_n}&\perp&z^{\ell_n-n+1},\dots, z^{\ell_n}\\
 r_{n-1}\propto z^{\ell_n-n+1} \wp_{n-1}^{\ell_{n}-1} &\perp& z^0,\dots, z^{n-2}
\eea
This  implies the following formal asymptotic data for $\GG_n$ near
all the singularities.\\

At $x=0$ we have
\bea
\GG_n(x)\sim G_n^{(0)}  \le[\matrix{1 & 0 \cr 0 & x^{n-1-\ell_n}{\rm
      e}^{V_{sing,0}(x)}}\ri]\le(\1 + \mathcal O(x)\ri) \\[20pt]
G_n^{(0)}:= \le[
\begin{array}{cc}
\frac {(-)^n
  \Delta_n^{\ell_n+1}}{\sqrt{\Delta_n^{\ell_n}\Delta_{n+1}^{\ell_{n+1}}}}
  &
-\frac {
  \Delta_{n+1}^{\ell_n}}{\sqrt{\Delta_n^{\ell_n}\Delta_{n+1}^{\ell_{n+1}}}}
\\[10pt]
\frac {
  \Delta_{n-1}^{\ell_n}}{\sqrt{\Delta_{n-1}^{\ell_{n-1}}\Delta_{n}^{\ell_{n}}}}
&
\frac {(-)^n
  \Delta_{n}^{\ell_n-1}}{\sqrt{\Delta_{n-1}^{\ell_{n-1}}\Delta_{n}^{\ell_{n}}}}
\end{array}
\ri]\ ,\ \ \det G_n^{(0)} = \frac 1{\Delta_{n+1}^{\ell_{n+1}}
      \Delta_{n-1}^{\ell_{n-1}}} \label{zeromon}
\eea
At $x=\infty$ we have
\bea
\GG_n(x) = G_n^{(\infty)} \le[\matrix {x^n & 0 \cr
0 & x^{-\ell_n-1} {\rm e}^{V_{sing,\infty}(x)}}\ri] \le(\1 + \mathcal O\le(\frac 1 x
      \ri)\ri)\\[10pt]
G_n^{(\infty)} = \le[\matrix {\frac 1 {\sqrt{h_n}} & 0\cr 0 &
      \sqrt{h_{n-1}}}\ri]
\label{inftymon}
\eea
Near any other pole $x=c$ of $V'(x)$ we have
\bea
\GG_n(x) = G_n^{(c)} \le[\matrix {1 & 0 \cr
0 & {\rm e}^{V_{sing,c}(x)} } \ri] \le(\1 + \mathcal O\le(x-c
      \ri)\ri)\\[10pt]
G_n^{(c)} := \le[\begin{array}{cc}
p_n(c) &\ds \sqrt{h_0}(Q-c)^{-1}_{n0} {\rm e}^{V_{reg,c}(c)} \\[10pt]
r_{n-1}^\star(c) &\ds   \sqrt{h_0}c^{n-\ell_n+1}(Q-c)^{-1}_{0n-1} {\rm e}^{V_{reg,c}(c)}
\end{array}\ri]
\eea
where in all these formul\ae\ the notation $V_{sing,p}$ ($V_{reg,p}$) denote the
singular (regular) part of $V$ at the point $p$.\par
Near a hard-edge point $x=a$ we have \cite{semiiso}
\bea
\GG_n \sim  G_n^{(a)} \le[\begin{array}{cc}
1 & \pm \varkappa \ln(x-a)\\
0 & 1
\end{array}\ri] \le(\1 + \mathcal O(x-a)\ri)\\[10pt]
G_n^{(a)} := \le[
\begin{array}{cc}
p_n(a) &  {\rm e}^{V(a)} \mathcal L\le(\frac {p_n(z)-p_n(a)}{a-z}\ri) \\
r_{n\!-\!1}^\star(a) & a^{n-1-\ell_n} {\rm e}^{V(a)} \mathcal L\le(\frac {r_n(z)-r_n(a)}{a-z}\ri)
\end{array}
\ri]
\eea
Together with these data we also have the jumps across the contours
$\Gamma_j$ defining our moment functional: the situation in this
respect  is identical to \cite{semiiso}.
In essence the matrix $\GG_n(x)$ has the following jumps across the contour $\Gamma_j$
\be
\GG_n(x)_+= \GG_n(x)_- \left[\begin{array}{cc}1 & 2i\pi\varkappa_i \\0 & 1\end{array}\right]\ .
\ee
Note that these jumps can be interpreted -depending on the point of
view- as the Stokes' multipliers of the problem near the
singularities.

\subsection{Isomonodromic Tau Function}
Using the results of \cite{semiiso} we find that the Jimbo-Miwa-Ueno isomonodromic tau
function \cite{JMU} is given by the same differential formul\ae\ in
Thm. \ref{cicciobello} {\bf provided} that we substitute the spectral
curve of the connection with the spectral curve of the connection in
the {\em traceless gauge}.
In our situation the trace of $\frac 1 x \DD_n(x)$ is  $V'(x) + \frac
{n-1-\ell_n}x$ so that we perform a scalar gauge transformation
\be
\mathcal A_n^{(JMU)} = \frac 1 x\DD_n(x) - \frac 12 \le(V'(x)+ \frac {n-1-\ell_n}x\ri)\1_{2\times 2}
\ee
This implies that the spectral parameter $y_{JMU}$ has the following
relation to the previously employed $y$;
\be
y_{JMU} = y + \frac 12\le(V'(x)+ \frac {n-1-\ell_n}x\ri)
\ee
This in turn implies that --up to multiplicative factors independent of
the isomonodromic times--
\bea
 \Delta_n^{\ell_n} &=&\mathcal F[V] \tau_{JMU}\\
&&\ln \mathcal F[V] = - \frac 12 \sum_{ c =\hbox{\small finite  pole of }\widehat V'}
\res{x=c}\widehat  V'_{sing,c}(x) \widehat V_{reg,c}(x)\\
&& \widehat V'(x) := V'(x) + \frac {n-1-\ell_n}x ,
\eea
where $\widehat V'_{sing,c}$ (  $\widehat V_{reg,c}$ ) denotes the singular
(regular) part of $\hat V'$ at the pole $c$.
\bx
\label{its}
For example let us consider the case relevant to the problem of the
probability of the longest increasing sequence of  random letter in a
word of fixed length \cite{ITW}
\bea
V(x) = -tx  - \sum_{\alpha=1}^M k_\alpha \ln\le(\frac {x-r_\alpha}{x} \ri)
\eea
In this case a direct computation ($\ell_n\equiv 0$) gives for
$\mathcal F$ the following expression
\bea
 \ln\mathcal F = -\frac {t}2 \sum_{\alpha=1}^{M} k_\alpha r_\alpha +
 \frac {n-1}2 \sum_{\alpha=1}^M k_\alpha\ln(-{r_\alpha}^2)
 + \frac 1 2 \sum_{\alpha=1}^M {k_\alpha}^2\ln (-{r_\alpha}^2)  - \frac 1 2 \sum_{\alpha=1}^{M}\sum_{\beta\neq \alpha}
\ln\le(\frac {r_\beta-r_\alpha}{r_\alpha r_\beta}\ri)^{ k_\alpha k_\beta }\label{nice}
\eea
which is the result obtained also in  formula (3.76)  \cite{ITW} :
note that in that formula $r_0=0$ and $k_0 = n-\sum_{\alpha=1}^{M}
r_\alpha$ and  a little of algebraic manipulation shows the
equivalence.  Moreover the signs inside the logarithms in (\ref{nice})
are in fact irrelevant since omitting them would amount to multiplying
$\mathcal F$ by a constant independent of the isomonodromic times, and
hence could be reabsorbed in  the definition of $\tau_{JMU}$.
\ex
\section{Schlesinger Transformations}
From the asymptotics that the shift $n\mapsto n+1$
implemented by the matrices $R_n$ are  --in the language of
isomonodromic deformations-- what is known as {\em elementary
  Schlesinger} transformations. Specifically the shift $n\mapsto n+1$
corresponds to the following two types of elementary Schlesinger
transformations according of the type of move ({\em circle} or {\em
  line}) (refer to formul\ae\ \ref{zeromon} and \ref{inftymon}):
\begin{enumerate}
\item [] {\bf Circle move}. The Schlesinger
  transformation adds one to the first entry of the formal monodromy
  at $\infty$ and subtracts one from  the second entry of the formal monodromy at zero
\item  []{\bf Line move}. The Schlesinger
  transformation adds one and subtracts one to the first and second
  entry (respectively) of the formal monodromy at infinity, leaving
  the formal monodromy at zero unchanged.
\end{enumerate}
However we can obtain a third type of elementary Schlesinger
transformation by considering two distinct sequences of LOPs
corresponding to two (weakly increasing) sequences of $\{\ell_n\}$'s.
Suppose indeed that we consider another sequence of LOPs and the
ensuing connection $x\pa_x - \widetilde{\DD_n}(x)$ for some fixed $n$ where the only
difference between the two pairs of LOPs is that one (or more) circle-moves have
been replaced by a line-move (or vice-versa) along the chain for
$n'\leq n$: the only difference in the formulas will be that
$\widetilde \ell_n = \ell_n\pm 1$. This
is implemented by the ``circle-to-line'' transformation $\mathcal T_n$
\ref{circleline} (suitably normalized).
\bea
&& \le[\matrix{p_n  \cr r^\star_{n-1} } \ri]=
\le[
\begin{array}{cc}
\ds a  + \frac b x
&
\ds \frac{c}{x}
\\[15pt]
\ds \frac {d}{x}
&
\ds \frac {e}{x}
\end{array}
\ri]\le[\matrix{\hat p_{n}\cr \hat r^\star_{n-1} }\ri]
\eea
where the coefficients $a,b,c,d,e$ above can be obtained explicitly in terms of shifted T\"oplitz determinants using the form of $\mathcal T_n$ (\ref{circleline}) and the normalizations (\ref{pirho}), and the polynomials $\hat p_n, \hat r^\star_{n-1}$ refer to the elements of the sequence of biorthogonal polynomials associated to the sequence $\{\hat \ell_k\}$: such sequence differs from $\{\ell_k\}$ because $\hat \ell_n = \ell_n-1$, namely there is a $k_0\leq n$ such that $\hat\ell_k = \ell_k-1,\ \forall k :\ \ k_0\leq k\leq n$.

 We therefore add the following third type of transformations;
\begin{enumerate}
\item []{\bf Circle-to-line move}. The Schlesinger
  transformation subtracts  one to the second entry of the formal monodromy
  at $\infty$ and adds one to the second entry  of the formal monodromy
  at zero.
\end{enumerate}

This last type of transformation shows that the orthogonal polynomials
on the line and the orthogonal polynomials on the circle are related
by a sequence $(n-1)$ Schlesinger transformations  and at each step
the Laurent biorthogonal polynomials that are obtained are those
appearing in the solution of integrable lattice hierarchies associated
to elementary orbits \cite{FG1}.
\section{Conclusion}
As a general ``philosophy'', it is acknowledged in the literature that
KP tau functions and isomonodromic tau functions are often, if not always,
related one to the other, in the sense that a KP (or Toda) tau function is an
isomonodromic tau function for a suitably chosen isomonodromic
deformation. In the case of orthogonal polynomials this relation was
explored in \cite{IKF} for some class and extended in
\cite{BEH,semiiso}. In this paper, this relation has been confirmed 
once more
for the particular generalized Toda systems associated to
``nonstandard'' minimal orbits of the Borel subgroup: the natural bridge
between the Hamiltonian and isomonodromic treatment is provided by the
solution of the inverse spectral problem in terms of bi-orthogonal
Laurent polynomials. It is to be expected that, whenever a
description or formulation of an integrable dynamical problem in terms
of (bi/multiple-orthogonal) polyomials is available, then a suitable
definition of the tau function for the associated isomonodromic
problem should tie the Hamiltonian tau function with the isomonodromic
one. For instance, in the case of the biorthogonal polynomials arising
in the study of two-matrix models \cite{B,BEH2} a natural
isomonodromic deformation of a polynomial connection can be derived;
however the connection is a  highly resonant one and at present a
definition of isomonodromic tau function for resonant deformations of
connections is not available. However it is possible to formulate such
a notion \cite{BM} and the connection can thus be positively be
established.\par

As it is recalled in the appendix to follow, the Laurent orthogonal
polynomials which we have investigated in the present paper are
related to the solution of the inverse spectral problem for Toda-like
systems associated to {\em certain} minimal (or elementary)
irreducible orbits. There exist in fact other minimal orbits for which
a treatment in terms of orthogonal polynomials of some sort is not
readily and generally available, although inspection of specific
examples leads to expect that it is possible to overcome the
difficulty. It is our intention to pursue the topic in future publications.

\section{Appendix: minimal irreducible co-adjoint orbits}

As it was mentioned earlier, every $n\times n$ principal submatrix of the Hessenberg matrix $Q$ that
defines recurrence relations (5-1) belongs to a $2n-2$-dimensional co-adjoint orbit of the Borel subgroup $\B_n$
of invertible upper triangular matrices in $sl(n)$. However, not every low-dimensional co-adjoint orbit can be obtained
this way. In this appendix, we give a description of all irreducible co-adjoint orbits of $\B_n$  in $sl(n)$ that have
a minimal dimension $2n-2$.

First, we introduce some notations. Let $\b_-$ be a subalgebra of lower triangular matrices  in $sl(n)$.
Denote by $J$ an $n\times n$ shift matrix (1s on the first superdiagonal and 0s everywhere else) and let
$Hess_n = J + \b_- $ denote a set of lower Hessenberg matrices. An element $Q\in Hess_n$ is called
{\bf reducible} if it  has a block upper triangular form $Q =\left [ \begin{array}{cc} Q_{11} & Q_{12}\\
0 & Q_{22}\end{array} \right ]$, where $Q_{11}$ is a $k\times k$ matrix ( $0 < k < n$ ). $Q$ is called
{\bf irreducible} otherwise.

Orbits of the co-adjoint action of $\B_n$ on $Hess_n$ are given by
\begin{equation}
\O_{Q_0}=\left \{ J + (\mbox{Ad}_b Q_0)_{\leq 0}\ : b\in \B_n\right \}\ .
\label{coad}
\end{equation}

It is easy to see that if $\O_{Q_0}$ contains a reducible (resp. irreducible) element then every element
of $\O_{Q_0}$ is reducible (resp. irreducible). Therefore it makes sense to talk about {\it irreducible orbits of
the co-adjoint action}. Our main goal in this appendix is to prove the following

\bt
An irreducible co-adjoint orbit of ${\mathbf B}_n$ in $Hess_n$ has a minimal dimension
$(2n-2)$ if and only if it contains an element $Q_0$ of the form
\be
{\displaystyle
Q_0=  J + H+ \sum_{\alpha=1}^k E_{i_\alpha,i_{\alpha-1}-\varepsilon_{\alpha-1}}
}
\label{normform}
\ee
where
\begin{enumerate}
\item $\varepsilon_{i} \in \{0,1\}$ and $\varepsilon_{0}=0$
\item $1=i_0 < i_1 -\varepsilon_1 \leq i_1 < i_2 -\varepsilon_{2} \leq i_2 < \cdots < i_{k-1} -\varepsilon_{k-1}
\leq i_{k-1} < i_k =n $
\item ${\displaystyle H=\sum_{\alpha \in \{1,\ldots, n\}\setminus \{i_0,\ldots, i_{k-1} \} } h_\alpha E_{\alpha\alpha}\ .}$
\end{enumerate}
\label{orbits}
\be
Q_0 = \le[\begin{array}{ccccccccccc}
0& 1 	& 	 &	  &	   &	          &	  &	&&&\\
 \vdots &\star 	&1	 &	  &	   &	          &	  &	&&&\\
 \vdots &         	& 0 &1	  &	   &	          &	  &	&&&\\
 1&	\cdots	&	  0  &\star	  &1	   &	          &	 &	&&&\\
  &		& \vdots	 &	0  &\star&1	          &	  &	&&&\\
  &		&  \vdots	 &	 \vdots&	   &\star  &1	  &	&&&\\
  &		&  \vdots	 &	\vdots  &	   &	 	 &\star&1	&&&\\
  &		&     1&0	  &	 \cdots  &	 \cdots	 &\cdots	  &0&1&&	    \\
  &&&&&&&\vdots &\star&1&\\
  &&&&&&&\vdots&&\star&1\\
  &&&&&&&1& 0  &\cdots &\star
\end{array}\ri]
\ee
[An example with $n=11$, $k=3$, $i_1 = 4, i_2=8,i_3=11$, $\epsilon_1=\epsilon_3=0$, $\epsilon_2=1$.]
\et

\br
The case of $H=h E_{nn}$ and $\varepsilon_\alpha=0  (\alpha =1,\ldots,
k-1)$ was studied in \cite{FG1,FG2}. It is orbits of this type that
can be studied via associated LOPs of the type appearing in this paper. Note that, in this case, parameters $\ell_j$ that were used in the main body of the paper are related to $i_\alpha$ via 
$$
\ell_j=\max\{i_\alpha: i_\alpha < j \}\ .  
$$
An investigation of propeties of
moment functionals connected with a more general minimal orbits
described in Theorem \ref{orbits} will appear elsewhere.
\er

Define a {\it staircase pattern} $(I,\varepsilon)$ as a collection of pairs of indices
\be
(I,\varepsilon)= \{ (i_1,1), (i_2,i_1 - \varepsilon_1), \ldots, (i_k=n, i_{k-1} - \varepsilon_{k-1}) \}\ ,
\label{stair}
\ee
where
$$
1=i_0 < i_1 -\varepsilon_1 \leq i_1 < i_2 -\varepsilon_{2} \leq i_2 < \cdots < i_{k-1} -\varepsilon_{k-1}
\leq i_{k-1} < i_k =n\ .
$$
In what follows we will often use a notation
$$
j_\alpha=i_{\alpha-1} - \varepsilon_{\alpha-1}\ .
$$
We say that $Q\in Hess_n$ has a staircase pattern $(I,\varepsilon)$ if
$$
Q_{i_\alpha, j_\alpha} \ne 0  \quad \mbox{and} \quad  Q_{ij}= 0 \quad \mbox{for}\quad  i>i_\alpha, \ j < j_{\alpha + 1}
\ (\alpha=1,\ldots, k)\ .
$$
The set of all matrices in $Hess_n$ that have a staircase pattern
$(I,\varepsilon)$ will be denoted by $Hess(I,\varepsilon)$. For
example, if $I=\{ 2,3,\ldots, n\}$ and $\varepsilon=\{ 0,0,\ldots,
0\}$, then $Hess(I,\varepsilon)$ coincides with the set of $n\times n$ Jacobi matrices.
An immediate property of the set $Hess(I,\varepsilon)$ is that it is stable under the co-adjoint
action of $\B_n$, since {\it corner entries} $Q_{i_\alpha, j_\alpha}$ and the entries "under the staircase"
$Q_{ij}= 0 \  i>i_\alpha, \ j < j_{\alpha + 1}$ have only zeroes to the left and below and, thus the former
are being acted upon only by the diagonal part of $\B_n$ and the latter cannot be made non-zero by the co-adjoint action.

Let us fix a staircase pattern $(I,\varepsilon)$.
To begin the proof of Theorem \ref{orbits}, we first employ the strategy used in \cite{gesha} to study {\it generic}
staircase orbits.
\bl
If $Q\in Hess(I,\varepsilon)$ then there exist $\tilde Q \in \O_Q$ such that
\bea
\nonumber
& \tilde Q_{i_\alpha j_\alpha} =1 \\
\nonumber
&\tilde Q_{i j_\alpha} =0 \quad (j_\alpha< i < i_\alpha )\\
\nonumber
&\tilde Q_{i_\alpha j} =0  \quad
(j_\alpha<  j < i_\alpha \quad  \mbox{and}\quad  j\ne j_\beta \ :\ \beta<\alpha\ ,\ j_\beta < i_\alpha )\\
& (\alpha = 1,\ldots, k )\ .
\label{reduced}
\eea

\label{reduce}
\el

{\bf Proof.} First, we use a diagonal conjugation to reduce $Q$ to an element with all corner
entries equal to 1 : $ Q \to Ad^*_{D} Q = D^{-1} (Q - J) D + J$, where $D=\mbox{diag}(d_1,\ldots, d_n)=
D_k \cdots D_1$ with diagonal matrices $D_\alpha$ defined by
$$
(D_\alpha)_{ii} = \left \{ \begin{array}{cc} 1 & i\ne i_\alpha\\
d_{i_\alpha}=\left (D_{\alpha-1}^{-1} \cdots D_{1}^{-1} Q D_1 \cdots D_{\alpha-1}\right )_{i_\alpha j_\alpha} & i=i_\alpha \end{array}
\right .
$$
Next, we use the co-adjoint action induced by a sequence of elementary upper-triangular matrices (each depending only on one parameter only)
to set as many as possible of the entries in rows and
columns occupied by corner entries equal to zero. More precisely, to eliminate an $(i, j_\alpha)$-entry
($j_\alpha \leq i < i_\alpha$) using
the corner entry $(i_\alpha, j_\alpha)$, one employs $Ad^*_{(\one + Q_{i j_\alpha} E_{i i_\alpha})}$.
Similarly, to eliminate an $(i_\alpha, j)$-entry ($j_\alpha < j < i_\alpha$),
one uses $Ad^*_{(\one - Q_{i_\alpha j} E_{j_\alpha j})}$. Note that when we write $Q_{i j_\alpha}$ (resp. $Q_{i_\alpha j}$),
we refer to entries of the "current" value of $Q$, i.e. to the element that belongs to the orbit through the initial $Q$
and that has been obtain through the sequence of transformations already applied.

The order in which we apply these elementary transformations is
defined as follows: we first set to zero the  entries in the 1st
column (going down the column), then in the $i_1$st row (moving right), then in the $j_2$nd column (moving down), then
in the $i_2$nd row (moving right) etc. Through the entire process, we want, for every $l < m$, to use an elementary
matrix of the from $\one + x E_{lm}$ at most once. This means, in particular, that any $(i_\alpha, j_\beta)$-entry, where
$\beta < \alpha$ and $j_\beta < i_\alpha$ cannot be touched, since a matrix of the form $\one + x E_{j_\beta i_\alpha}$
has already been used to eliminate the $(j_\beta, j_\alpha)$-entry. This explains why entries
$\{ (i_\alpha, j_\beta) : \beta < \alpha, j_\beta < i_\alpha \}$ are excluded from the list of entries in (\ref{reduced}).
On the other hand, all non-corner entries that are in the list can be set to 0, regardless of their initial values.
Q.E.D.

\begin{corollary}
For each  $Q\in Hess(I,\varepsilon)$  the matrix entries specified in
(\ref{reduced}) are independent functions on $\O_Q$.
\end{corollary}

{\bf Proof.} It suffices to notice that applying to $\tilde Q$ constructed in Lemma \ref{reduce} elementary
transformations of the same type that was used in its construction, but in the reverse order and with arbitrary
parameters, one can obtain an element in $\O_Q$ with arbitrary nonzero values of the corner entries and arbitrary
values of non-corner values specified in (\ref{reduced}).
Q.E.D.

\bl
If, for some $1\leq \alpha < k$, $\varepsilon_\alpha > 1$, then, for any $Q\in Hess(I,\varepsilon)$,
$\dim \O_Q > 2n-2$.
\label{count}
\el

{\bf Proof.} Denote by $M(I,\varepsilon)$ the set of pairs of indices that appear in the list given
in (\ref{reduced}). In view of the corollary above, we only need to show that, under conditions of the Lemma, the
number of elements in $M(I,\varepsilon)$ is greater than $2n-2$. We will also show that, if $0\leq \varepsilon_\alpha \leq 1$
for $\alpha=1,\ldots, k-1$, then $\#M(I,\varepsilon)=2n - 2$.

We will use an induction on $k$ and $n$. Clearly, if $k=1$ then $\varepsilon_0=0$ and $\#M(I,\varepsilon)=2n - 2$.
Moreover, $\#M(I,\varepsilon)=2n - 2$ for any $k$, provided $\varepsilon_\alpha=0$ for all $\alpha$. Let now $k=2$ and
$\epsilon_1 > 0$. We are looking for a number of elements in the set
$\{(1,1),\ldots, (i_1,1), (i_1,2),\ldots , (i_1, i_1 -1); (i_1-\varepsilon_1, i_1-\varepsilon_1), \ldots,
(i_1-1,i_1-\varepsilon_1, i_1-\varepsilon_1), (i_1+1,i_1-\varepsilon_1, i_1-\varepsilon_1), \ldots, (n, i_1-\varepsilon_1), \ldots
 (n, i_1-1), (n, i_1 + 1), \ldots, (n, n-1)\}$, which is equal to $2(i_1-1) + 2(n-(i_1-\varepsilon_1)) - 2 =
 2(n +\varepsilon_1 - 2) \  \left \{ \begin{array}{cc} =2n -2 &\quad  \mbox{if}\quad  \varepsilon_1 =1\\
 > 2n - 2 & \quad \mbox{if}\quad  \varepsilon_1 >1 \end{array} \right . $.

For $k>2$, let $s$ be such that $j_s < i_1 \leq j_{s+1}$. We first consider the case when there is no $r$ such that $j_r=i_1$. Then the set
$M(I,\varepsilon) \setminus \{(1,1),\ldots, (i_1,1), (i_1,2),\ldots , (i_1, i_1 -1)\}$
has the same cardinality as a set $M(I',\varepsilon')$, where\\
$
(I',\varepsilon')=\left \{(i_2 -j_2,1), (i_3-j_2, j_3-j_2 +1), \ldots,  (i_s-j_2, j_s -j_2+1), (i_{s+1} - j_2, j_{s+1} - j_2), \ldots,
(i_k -j_2, j_k - j_2)  \right \}
=\left \{ (i_2-i_1+ \varepsilon_1,1), (i_3 -i_1 + \varepsilon_1,i_2 - i_1 + \varepsilon_1 - (\varepsilon_2 -1)), \ldots,
 (i_s -i_1 + \varepsilon_1,i_{s-1} - i_1 + \varepsilon_1 - (\varepsilon_{s-1} -1)),\right .$\\
 $\left . (i_{s+1} -i_1 + \varepsilon_1,i_s - i_1 + \varepsilon_1 - \varepsilon_s),\ldots,
 (i_k -i_1 + \varepsilon_1,i_{k-1} - i_1 + \varepsilon_1 - \varepsilon_{k-1}) \right \}
$, that is
$$ n'= i_{k-1}'= i_k -i_1 + \varepsilon_1\ ,$$
$$ i_{\alpha}'= i_{\alpha+1}  -i_1 + \varepsilon_1\quad \alpha=1,\ldots, k-1 $$
and
$$ \varepsilon'_\alpha = \varepsilon_{\alpha+1} - 1 \quad ( 1\leq \alpha\leq s-2 )\ , \quad
 \varepsilon'_\alpha = \varepsilon_{\alpha+1} \quad ( s-1\leq \alpha\leq k-2 )\ .
$$
If $s\geq 2$ then $\varepsilon_1 \geq 1$ and  $\varepsilon_s=i_s-j_s > i_s - i_1 \geq s-1 \geq 1$, so that
$\varepsilon'_{s-1}=\varepsilon_s \geq 2$ and, by the induction hypothesis,
$\#M(I',\varepsilon')> 2 ( n-i_1 + \varepsilon_1 - 1) \geq 2(n-i_1)$ and
$\#M(I,\varepsilon)> 2(i_1 -1) + 2 ( n-i_1)= 2(n-1)$.

If $s=1$  then $\varepsilon'_\alpha=\varepsilon_{\alpha +1}$ for $ 1 < \alpha \leq {k-2}$ and
$$
\#M(I',\varepsilon') \left \{ \begin{array}{cc} =2(n - i_1 + \varepsilon_1 -1)  &\quad  \mbox{if all} \quad
\varepsilon'_\alpha \leq 1\\
 > 2(n - i_1 + \varepsilon_1 -1) & \quad \mbox{if some}\quad  \varepsilon'_\alpha >1 \end{array} \right .
$$
and thus, $\#M(I,\varepsilon)= 2 (i_1-1) + \#M(I',\varepsilon')$ is greater than $2n-2$ if $\varepsilon_\alpha >1$ for
some $\alpha>1$ and is equal to $2n-2$ otherwise.

Finally, consider the case when $j_r=i_1$ for some $r>1$. If $r>2$ then $\varepsilon_r=i_r-j_r=i_r-i_1\geq r-1\geq 2.$
Define $(\tilde I, \tilde \varepsilon) = (I,\varepsilon) \setminus \{ (i_2, j_2), \ldots, (i_{r-1}, j_{r-1})\}$.
Then $(\tilde I, \tilde \varepsilon)$ still defines an irreducible staircase pattern, $\tilde k = \# (\tilde I, \tilde \varepsilon)
, k$ and $\#M(I,\varepsilon) > \#M(\tilde I,\tilde \varepsilon) > 2n-2$ by the induction hypothesis.

If $r=2$ then $\varepsilon_1=0$, $j_2=i_1$ and $\#M(I,\varepsilon)= 2 (i_1-1) + \#M(I',\varepsilon')$, where
$(I', \varepsilon') = \{ (i_2-i_1+1, 1), (i_3-i_1+1, i_2-i_1+1 -\epsilon_2),\ldots, (n-i_1+1,
(i_{k-1}-i_1+1 -\epsilon_{k-1}) \}$ and, again by induction, the statement follows.
Q.E.D.

We are now ready to complete the proof of Thm. \ref{orbits}.
\medskip

{\bf Proof of Theorem \ref{orbits}.} Assume that $\dim \O_Q=2n-2$. We have shown that if $Q \in Hess(I, \epsilon)$
then $\epsilon_\alpha \leq 1$ for $\alpha=1, \ldots, k-1$. Assume that the latter condition is satisfied
and consider the element $\tilde Q$ constructed in Lemma \ref{reduce}. Suppose that some non-corner
entry $\tilde Q_{ij}$ $(i>j)$ is nonzero. Then, by construction of $\tilde Q$, $j\ne j_\alpha$
($\alpha=1, \ldots, k$).  Define a diagonal matrix $D$ by
$$
D_{ll}= \left \{
\begin{array}{cc} 1 & \quad \mbox{if} \quad l\ne j\quad \mbox{or} \quad l\ne j_\beta\quad \mbox{or}\quad  i\ne i_\beta\\
d & \quad\quad \quad \mbox{if}\quad l=j \quad \mbox{or} \quad (\ l=j_\beta\quad \mbox{and}\quad i=i_\beta\ )
\end{array}
\right .\ .
$$
Then $\mbox{Ad}^*_D \tilde Q = D^{-1} \tilde Q D$ has the same values
as $\tilde Q$ in the  entries specified by
(\ref{reduced}) but $(\mbox{Ad}^*_D \tilde Q)_{ij} =d^{-1} \tilde Q_{ij}$. This means that the matrix entry
$Q_{ij}$ viewed as a function on $\O_Q$ is independent of the  matrix entries specified by
(\ref{reduced}), which is in contradiction with  $\dim \O_Q = 2n-2$. Therefore, $\tilde Q_{ij}=0$ for
all $(i,j) \ne (i_\alpha, j\alpha)$. Since, by Lemma \ref{reduce}, $\tilde Q_{j_\alpha j_\alpha} =0$
for $\alpha = 1, \ldots, k$, we proved that $\dim \O_Q=2n-2$ implies that $\O_Q$ contains an element of the form
(\ref{normform}).

To prove the converse consider an element  $Q_0$ defined by (\ref{normform}).
Clearly, for any $b\in \B_n$, $\mbox{Ad}^*_b Q_0= \mbox{Ad}^*_b(Q_0 - H) + H$, therefore it is sufficient to consider
the case where $h=0$. In other words, we are interested in parametrizing the set
$$
\left \{(b\ ( Q_0 - H - J)\ b^{-1})_{\leq 0}\ : b\in \B_n\right \}\ .
$$
Note that, for $i>j$ we have
$$(b\ E_{ij}\ b^{-1})_{\leq 0} = ((b\ e_{i}) (e^T_j\ b^{-1}))_{\leq 0} = (u v^T)_{\leq 0}\ ,$$
where
$$
u=(\Pi_i -\Pi_{j-1}) (b\ e_{i}), \ v^T = (e^T_j\ b^{-1}) (\Pi_i -\Pi_{j-1})\ .
$$
Thus,
\be
(b\ ( Q_0 - H - J)\ b^{-1})_{\leq 0} = \sum_{\alpha=1}^k (u_\alpha v_\alpha^T)_{\leq 0}
\label{prmtrztn}
\ee
with
$$
u_\alpha=(\Pi_{i_\alpha} -\Pi_{i_{\alpha-1}-\varepsilon_{\alpha-1} - 1}) (b\ e_{i_\alpha}),
\ v_\alpha^T = (e^T_{i_{\alpha-1}-\varepsilon_{\alpha-1}}\ b^{-1})
(\Pi_{i_\alpha} -\Pi_{i_{\alpha-1}-\varepsilon_{\alpha-1} - 1}) \ .
$$
Entries of vectors $u_\alpha, v_\alpha$ cannot be arbitrary. First,
\be
v^T_\alpha u_\alpha = e^T_{j_{\alpha}}\ b^{-1} (\Pi_{i_\alpha} -\Pi_{j_{\alpha} - 1})) b e_{i_\alpha} =
e^T_{j_{\alpha}} e_{i_\alpha} =0\ .
\label{restr1}
\ee
Next, if $\varepsilon_\alpha=0$, i.e. $j_\alpha=i_{\alpha-1}$, then
\be
(v_\alpha)_{j_{\alpha}}=(b^{-1})_{j_{\alpha} j_{\alpha}}= (u_{\alpha-1})^{-1}_{j_{\alpha}} \ .
\label{restr2}
\ee
Finally, if $\varepsilon_\alpha=1$, i.e. $j_\alpha=i_{\alpha-1}-1$, then
\be
v^T_\alpha u_{\alpha-1} =(v_\alpha)_{j_{\alpha}} (u_{\alpha-1})_{i_{\alpha-1}-1} +
(v_\alpha)_{j_{\alpha}+1} (u_{\alpha-1})_{i_{\alpha-1}} = (b^{-1})_{j_{\alpha} j_{\alpha}} b_{j_{\alpha} j_{\alpha}+1} +
(b^{-1})_{j_{\alpha} j_{\alpha}+1} b_{j_{\alpha}+1 j_{\alpha}+1}=0
\ .
\label{restr3}
\ee
We claim that (\ref{restr1}),(\ref{restr2}),(\ref{restr3}) are the only restrictions on $u_\alpha, v_\alpha$.
We will verify this claim for $k=2$. The general case follows by an easy induction.

If $\epsilon_1=0$, we set $u_1=\mbox{col} [u_{11}, u_{12}, u_{13},0,\ldots, 0 ]$  and
$v^T_1=[v_{11}, v^T_{12}, v_{13},0,\ldots, 0 ]$, where $u_{11}, u_{13}\ne 0, v_{11}\ne 0, v_{13} \in \C$
and $u_{12},  v_{12} \in \C^{i_1-2}$. Similarly,
$u_2=\mbox{col} [0,\ldots, 0, u_{21}, u_{22}, u_{23} ]$ and
$v^T_2=[0,\ldots, 0, v_{21}=u_{13}^{-1}, v^T_{22}, v_{23} ]$, where $u_{21}, u_{23}\ne 0,  v_{23} \in \C$
and $u_{22},  v_{22} \in \C^{n-i_1-1}$. We assume that conditions (\ref{restr1}) are satisfied: $v_1^T u_1=v_2^T u_2=0$
and define
$$
b = \left (
\begin{array}{ccccc}
v_{11}^{-1} &  - v_{11}^{-1}v_{12}^{T} & u_{11} & 0 & 0 \\
0 & \one & u_{12} & 0 & 0\\
0 & 0 & u_{13} & -u_{13} v_{22}^{T} & u_{21}\\
0 & 0 & 0 & \one & u_{22}\\
0 & 0 & 0 & 0 & u_{23}
\end{array}
\right ) \quad , \quad
b^{-1} = \left (
\begin{array}{ccccc}
v_{11} &  v_{12}^{T} & v_{13} & * & * \\
0 & \one & -u_{12}u_{13}^{-1} & * & * \\
0 & 0 & v_{21} &  v_{22}^{T} & v_{23}\\
0 & 0 & 0 & \one & -u_{22}u_{23}^{-1}\\
0 & 0 & 0 & 0 & u_{23}^{-1}
\end{array}
\right )
\ .
$$
The specified entries are consistent with the relation $b b^{-1} = \one$ and entries marked
by $*$s are uniquely determined by this relation.

Similarly, if $\epsilon_1=1$, we set $u_1=\mbox{col} [u_{11}, u_{12}, u_{13},u_{14},0,\ldots, 0 ]$  and
$v^T_1=[v_{11}, v^T_{12}, v_{13},v_{14}, 0,\ldots, 0 ]$, where $u_{11}, u_{13}, u_{14}\ne 0, v_{11}\ne 0, v_{13}, v_{14} \in \C$
and $u_{12},  v_{12} \in \C^{i_1-3}$; and
$u_2=\mbox{col} [0,\ldots, 0, u_{21}, u_{22}, u_{23}, u_{24} ]$ and
$v^T_2=[0,\ldots, 0, v_{21}, v_{22}=-v_{21}\frac{u_{13}}{u_{14}}, v^T_{23}, v_{24} ]$, where $u_{21}, u_{22}, u_{24}\ne 0,
v_{21}\ne 0, v_{24} \in \C$
and $u_{23},  v_{23} \in \C^{n-i_1-2}$. Assuming again that  $v_1^T u_1=v_2^T u_2=0$,
define
$${\small
b = \left (
\begin{array}{cccccc}
v_{11}^{-1} &  - v_{11}^{-1}v_{12}^{T} & - v_{11}^{-1}v_{13} & u_{11} & 0 & 0 \\
0 & \one & 0 & u_{12} & 0 & 0\\
0 & 0 & v_{21}^{-1} & u_{13} & - v_{21}^{-1}v_{23}^{T} & u_{21}\\
0 & 0 & 0 & u_{14} & 0 & u_{22}\\
0 & 0 & 0 & 0 & \one & u_{23}\\
0 & 0 & 0 & 0 & 0 & u_{24}
\end{array}
\right ) \quad , \quad
b^{-1} = \left (
\begin{array}{cccccc}
v_{11} &  v_{12}^{T} & v_{13} & v_{14} & * & * \\
0 & \one & 0 &  -u_{12}u_{14}^{-1} & * & * \\
0 & 0 & v_{21} & v_{22} &  v_{23}^{T} & v_{24}\\
0 & 0 & 0 & u_{14}^{-1} & 0 & -u_{22} u_{24}^{-1}\\
0 & 0 & 0 & 0& \one & -u_{23}u_{24}^{-1}\\
0 & 0 & 0 & 0 & 0 & u_{24}^{-1}
\end{array}
\right )
}
$$
and observe that (\ref{restr1}), (\ref{restr3}) are consistent with $b b^{-1} = \one$ and entries marked
by $*$s can be  uniquely determined .

To conclude the proof, observe that the right hand side of
(\ref{prmtrztn}) is invariant under a transformation
$ u_\alpha \to t_\alpha u_\alpha,\ v_\alpha \to t^{-1}_\alpha v_\alpha$, where $t_\alpha$ are arbitrary non-zero
parameters. Therefore, we can assume that
\be
(v_\alpha)_{j_\alpha}= 1\quad \mbox{if} \quad \alpha=1 \quad \mbox{or}
\quad \varepsilon_{\alpha-1}=1 \label{norm1}
\ee
Recall, that, if $\varepsilon_{\alpha-1}=0$, then $(v_\alpha)_{j_\alpha}$ is given by
(\ref{restr2}), while $(u_\alpha)_{j_\alpha}$ is determined by the condition (\ref{restr1}) for all $\alpha$.
Furthermore, if $\varepsilon_{\alpha-1}=1$, then $j_{\alpha}=i_{\alpha-1} +1$ and $(j_\alpha,j_\alpha)$, $(j_\alpha+1,j_\alpha)$ and
$(j_\alpha+1,j_\alpha+1)$-entries of the right hand side of (\ref{prmtrztn}) are given by
\be
\begin{array}{cc}
(j_\alpha,j_\alpha)\ : & (u_{\alpha-1})_{j_\alpha} (v_{\alpha-1})_{j_\alpha}  + (u_{\alpha})_{j_\alpha+1}
\frac{(u_{\alpha-1})_{j_\alpha}}{(u_{\alpha-1})_{j_\alpha+1}} - \sum_{s=j_\alpha+2}^{i_\alpha} (u_\alpha)_s (v_\alpha)_s\\
(j_\alpha + 1,j_\alpha)\ : & (u_{\alpha-1})_{j_\alpha+1} (v_{\alpha-1})_{j_\alpha}  + (u_{\alpha})_{j_\alpha+1}\\
(j_\alpha + 1,j_\alpha + 1)\ : & (u_{\alpha-1})_{j_\alpha+1} (v_{\alpha-1})_{j_\alpha+1} -  (u_{\alpha})_{j_\alpha+1}
\frac{(u_{\alpha-1})_{j_\alpha}}{(u_{\alpha-1})_{j_\alpha+1}}\ ,
\end{array}
\ee
where we have used (\ref{restr1}), (\ref{restr3}) and (13-9). Note that the entries in (13-10) are the only entries
in (\ref{prmtrztn}) that depend on $(v_{\alpha-1})_{j_\alpha},(v_{\alpha-1})_{j_\alpha}$ and $(u_{\alpha})_{j_\alpha+1}$.
Moreover, (13-10) does not change under a transformation
$$
(u_{\alpha})_{j_\alpha+1} \to (u_{\alpha})_{j_\alpha+1} - t (u_{\alpha-1})_{j_\alpha+1}\ , \
(v_{\alpha-1})_{j_\alpha} \to (v_{\alpha-1})_{j_\alpha} + t\ , \
(v_{\alpha-1})_{j_\alpha+1} \to (v_{\alpha-1})_{j_\alpha} - t \frac{(u_{\alpha-1})_{j_\alpha}}{(u_{\alpha-1})_{j_\alpha+1}}\ .
$$
This means that we can set
\be
(u_{\alpha})_{j_\alpha+1}=(u_{\alpha})_{i_{\alpha-1}}=0 \quad \mbox{or} \quad \varepsilon_{\alpha-1}=1\ \label{norm2}
\ee
Under the normalizations (\ref{norm1}), (\ref{norm2}) and restrictions (\ref{restr1}),(\ref{restr2}),(\ref{restr3}), the rest
of the parameters in (\ref{prmtrztn}),
$$
(u_\alpha)_s, (v_\alpha)_s\ ,\ s=i_{\alpha-1}+1,\ldots, i_\alpha\ ,\ \alpha=1,\ldots, k\ ,
$$
can be chosen arbitrarily and, on the other hand, these parameters are uniquely determined by
the right hand side of (\ref{prmtrztn}). Thus, for $Q_0$ satisfying conditions
of Theorem \ref{orbits} we have found an explicit parametrization of $\O_{Q_0}$ by $2n-2$
independent parameters, which completes the proof.
Q.E.D.

\end{document}